\documentclass[twocolumn,superscriptaddress,amsmath,amssymb,aps,prb,notitlepage,floatfix,longbibliography,10pt]{revtex4-1}

\usepackage{graphicx}
\usepackage{dcolumn}
\usepackage{bm}
\usepackage{physics}
\usepackage[colorlinks=true,linkcolor=red]{hyperref}
\usepackage{xcolor}
\usepackage{xfrac} 
\usepackage[normalem]{ulem}

\newcommand{\pcsadd}{Center for Theoretical Physics of Complex Systems, Institute for Basic Science(IBS), Daejeon 34126, Korea}
\newcommand{\ustadd}{Basic Science Program(IBS School), Korea University of Science and Technology(UST), Daejeon 34113, Korea}

\newcommand{\mbk}{\mathbf{k}}

\newcommand{\vpsi}{\vec{\psi}}
\newcommand{\bpsi}[1]{\bra{\psi_{#1}}}
\newcommand{\kpsi}[1]{\ket{\psi_{#1}}}

\newcommand{\kx}{\ket{x}}
\newcommand{\bx}{\bra{x}}
\newcommand{\ky}{\ket{y}}
\newcommand{\by}{\bra{y}}

\newcommand{\efb}{E_\text{FB}}
\newcommand{\mus}{\mathbf{U}}

\begin{document}

\title{Flatband generator in two dimensions}

\author{Wulayimu Maimaiti}
\affiliation{\pcsadd}
\affiliation{\ustadd}
\affiliation{Department of Physics and Astronomy, Center for Materials Theory, Rutgers University, Piscataway, NJ 08854, USA}

\author{Alexei Andreanov}
\affiliation{\pcsadd}
\affiliation{\ustadd}

\author{Sergej Flach}
\affiliation{\pcsadd}
\affiliation{\ustadd}

\date{\today}

\begin{abstract}
    Dispersionless bands -- \emph{flatbands} -- provide an excellent testbed for novel physical phases due to the fine-tuned character of flatband tight-binding Hamiltonians. The accompanying macroscopic degeneracy makes any perturbation relevant, no matter how small. For short-range hoppings flatbands support compact localized states, which allowed to develop systematic flatband generators in $d=1$ dimension in Phys. Rev. B {\bf 95} 115135 (2017) and Phys. Rev. B {\bf 99} 125129 (2019). Here we extend this generator approach to $d=2$ dimensions. The \emph{shape} of a compact localized state turns into an important additional flatband classifier. This allows us to obtain analytical solutions for classes of $d=2$ flatband networks and to re-classify and re-obtain known ones, such as the checkerboard, kagome, Lieb and Tasaki lattices. Our generator can be straightforwardly generalized to three lattice dimensions as well.
\end{abstract}

\maketitle

\section{Introduction}
\label{sec:introduction}

Physical systems with macroscopic degeneracies have attracted a lot of attention during the last decades. Such degeneracies are highly sensitive to even slightest perturbations which makes them perfect testbeds for identifying and  studying various exotic or unconventional correlated phases of matter. Flatbands are dispersionless energy bands of translationally invariant tight-binding networks.~\cite{derzhko2015strongly,leykam2018artificial,maimaiti2020thesis} The absence of dispersion implies a macroscopic degeneracy of the flatband eigenstates. A flatband (FB) results from destructive interference of the hoppings which requires their fine-tuning. All the known FB examples with short-range hopping support compact localized states (CLS)~\cite{read2017compactly} as eigenstates with strictly compact support. FB networks were extensively studied theoretically in $d=1$,~\cite{derzhko2006universal,derzhko2010low,hyrkas2013many} $d=2$~\cite{mielke1991ferromagnetism,tasaki1992ferromagnetism,misumi2017new} and  $d=3$~\cite{mielke1991ferromagnetism,nishino2005three,lieb1989two,mielke1991ferromagnetic,mielke1992exact,brandt1992hubbard,ramachandran2017chiral} lattice dimensions. Models featuring FBs have been experimentally realized in a variety of settings, including optical wave guide networks,~\cite{zong2016observation,mukherjee2017observation,mukherjee2015observation1,vicencio2015observation,mukherjee2015observation} exciton-polariton condensates,~\cite{masumoto2012exciton,jacqmin2014direct,biondi2015incompressible,klembt2017polariton} and ultra-cold atomic condensates.~\cite{guzman2014experimental,vicencio2015observation,mukherjee2015observation1,weimann2016transport,xia2016demonstration,taie2015coherent,jo2012ultracold,masumoto2012exciton,baboux2016bosonic}  

Due to their fine-tuned character, flatband systems are fragile to perturbations that can easily destroy the macroscopic degeneracy. As a consequence, exotic phases of matter with unusual properties emerge under the effect of various perturbations: disorder,~\cite{vidal2001disorder,chalker2010anderson,leykam2017localization,bodyfelt2014flatbands,danieli2015flatband} external fields,~\cite{khomeriki2016landau,kolovsky2018topological,rhim2020quantum}nonlinearities.~\cite{danieli2018compact,johansson2015compactification,real2018controlled} Interactions in FB lead to a plethora of interesting phenomena: delocalisation and conserved quantities,~\cite{vidal2000interaction,doucot2002pairing,tovmasyan2018preformed,danieli2020quantum} disorder free many-body  localisation,~\cite{tovmasyan2018preformed,danieli2020many,kuno2020flat,orito2020exact} groundstate ferromagnetism,~\cite{mielke1991ferromagnetic,mielke1991ferromagnetism,tasaki1992ferromagnetism,mielke1993ferromagnetism,tasaki2008hubbard,tasaki1994stability,maksymenko2012flatband}  pair formation for hard core bosons,~\cite{mielke2018pair} superfluidity,~\cite{peotta2015superfluidity,julku2016geometric,mondaini2018pairing} and superconductivity.~\cite{volovik2018graphite}

However the very defining feature of the flatbands -- their fine-tuned degeneracy -- makes it difficult to identify them in the relevant Hamiltonian parameter space. A number of methods has been proposed to construct FB lattices: line graph approach,~\cite{mielke1991ferromagnetism}  extended cell construction method,~\cite{tasaki1992ferromagnetism} origami rules,~\cite{dias2015origami} repetitions of mini-arrays,~\cite{morales2016simple} local symmetry partitioning,~\cite{roentgen2018compact} chiral symmetry,~\cite{ramachandran2017chiral}, fine-tuning relying on specific CLS and network symmetries,~\cite{nishino2003flat,nishino2005three} using specific properties of FB,~\cite{rhim2019classification} \textit{etc}. All these methods apply to either specific geometries of the underlying networks or to networks with particular symmetries. 

All the above discussed FB networks support CLS. It follows that the properties of the CLS together with a number of generic network properties form a set of classifiers which will fix a particular FB network class. That approach leads to systematic FB generators based on these classifiers. The CLS classifiers are its size $U$ (of occupied unit cells) and shape (in dimensions $d \geq 2$), while the generic network classifiers are its dimension $d$, the Bravais lattice type, the hopping range, and the number of bands $\nu$ (i.e. the number of sites per unit cell). The simplest generator case $U=1$ with arbitrary remaining network classifiers was obtained in Ref.~\onlinecite{flach2014detangling}. The more sophisticated case $U=2$ with $d=1$, nearest neighbour unit cell hopping and two bands $\nu=2$ was solved in closed form in  Ref.~\onlinecite{maimaiti2017compact}, with its extension to larger band number $\nu$ and CLS size $U$ published in Ref.~\onlinecite{maimaiti2019universal}.

In this work, we extend the $d=1$ FB generator~\cite{maimaiti2017compact,maimaiti2019universal} to two dimensions $d=2$ and indicate the road to generators in dimension $d=3$. We introduce a systematic classification of $d=2$ FB networks using their CLS classifiers - size $U$ and shape. We demonstrate how to find analytic solutions for some FB classes. We re-generate and classify some of the already known $d=2$ FB lattices such as the checkerboard, kagome, Lieb, and Tasaki lattices, along with a multitude of completely new $d=2$ FB lattices. 

The paper is organized as follows. In Section~\ref{sec:definitions}, we introduce the main definitions and conventions that we use throught the text. These conventions are direct generalization to $d=2$ of the conventions used for the $d=1$ FB generator.~\cite{maimaiti2017compact,maimaiti2019universal} Section~\ref{sec:2d-cls-class} introduces CLS of $d=2$ FB Hamiltonians, and the classification of FB Hamiltonians by their CLS. It also presents an inverse eigenvalue problem for finding Hamiltonians from a given CLS class. An algorithm to solve these inverse eigenvalue problems is turned into an efficient FB generator in Sec.~\ref{sec:2d-fb-gen}. In the following Sec.~\ref{sec:results}, we apply the FB generator to some classes of CLS, illustrate how the inverse eigenvalue problem is resolved and show the results. We conclude by summarising our results and discussing open problems.

\section{Main definitions}
\label{sec:definitions}

\subsection{Models}

\begin{figure}
    \centering
    \includegraphics[width=0.9\columnwidth]{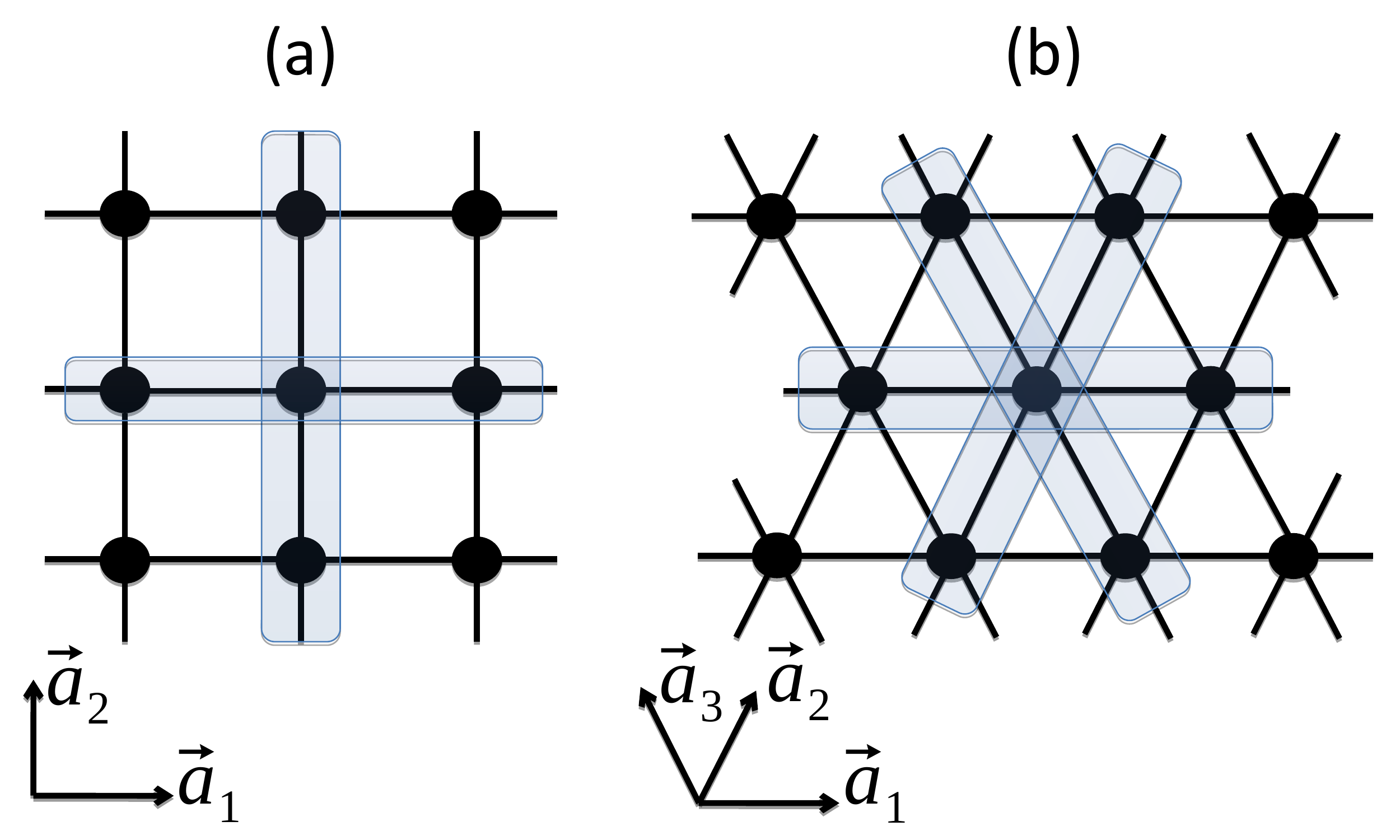}
    \caption{Example of nearest neighbor unit cells for 2D lattices, where $\vec{a}_1,\ \vec{a}_2$ are primitive translation vectors and $\vec{a}_3 = \vec{a}_1 + \vec{a}_2$. In our conventions: (a) a square lattice has only $2$ n.n. hopping directions: $\vec{a}_1,\ \vec{a}_2$, (b) a triangular/hexagonal lattice has $3$ n.n. hopping directions: $\vec{a}_{i=1,2,3}$.}
    \label{fig:square-vs-triangular}
\end{figure}

We consider a $d=2$ translational invariant tight-binding lattice with $\nu$ sites per unit cell. We use the same notation as in our previous works, Ref.~\onlinecite{maimaiti2017compact,maimaiti2019universal}: we label wavefunctions by the unit cell index $n$, so that the full wavefunction reads $\Psi = (\vpsi_1, \dots, \vpsi_n, \dots)$, where the $\nu$-component vector $\vpsi_n$ is the wave function of the $n$th unit cell. For simplicity we restrict to nonzero hopping between nearest neighbour unit cells only, and note that the generalization to longer range hoppings is cumbersome but straightforward. Nearest neighbor unit cells are defined along (combinations of) primitive lattice translation vectors. A set of matrices $H_{\chi}$ describes the hopping between different pairs of unit cells. The index $\chi$ encodes the direction of the hopping - $x,y$,\emph{etc}. The result is illustrated in Fig.~\ref{fig:square-vs-triangular} for a square lattice and a triangular lattice, the only possible cases for the n.n. directions in $d=2$. A unit cell on the square lattice has $4$ neighbours and $2$ different directions of n.n. hopping - both along the primitive lattice translations vectors $\vec{a}_1$ and $\vec{a}_2$. A site on a triangular lattice has $6$ neighbours and $3$ possible directions of n.n. hopping: $\vec{a}_{i=1,2,3}$ only two of which corresponds to the primitive lattice translation vectors. With these conventions the Hamiltonian eigenvalue problem reads
\begin{gather}
    \label{eq:2d-nn-mat-rep}
    H_0 \vpsi_n + \sum_{\chi}\left(H_{\chi}^\dagger \vpsi_{n_{\chi}'} + H_{\chi} \vpsi_{n_{\chi}}\right) = E \vpsi_n, \quad n\in \mathbb{Z} \;.
\end{gather} 
Here $H_0$ describes the intracell hopping and $H_{\chi}$ is the nearest neighbor hopping matrix for the $\chi$th direction, with $n_{\chi}$ and $n_{\chi}'$ being the respective indices of the two nearest neighboring unit cells along the $\chi$th direction. Because of the translation invariance the Floquet-Bloch theorem applies and the eigenstates of Eq.~\eqref{eq:2d-nn-mat-rep} can be expressed as $\vpsi_n = \vec{u}(\mbk) e^{-i \mbk \cdot \vec{R}_n}$, where the Bloch polarization vector $\vec{u}(\mbk)$ has $\nu$ components ${u}_{\mu}(\mbk), \mu=1,\dots,\nu$. Finally $\mbk=(k_x, k_y)$ is the wave vector, and $\vec{R}_n$ is the position of the $n$th unit cell. Then the eigenvalue problem~\eqref{eq:2d-nn-mat-rep} in momentum space reads
\begin{gather}
	H_{\mbk} \vec{u}(\mbk) = E(\mbk) \vec{u}(\mbk).
\end{gather}
The eigenvalues $E(\mbk)$ provide the band structure of the Hamiltonian $H_{\mbk}$.

\subsection{Classification of compact localised states}
\label{sec:2d-cls-class}

In our previous work~\cite{maimaiti2019universal} we have shown that the size $U$ of a CLS is the only CLS-related flatband classifier in dimension $d=1$ dimension. For $d=2$ dimensions, size and shape of the CLS plaquette turn into relevant flatband classifiers. The size of a CLS is given by two integers $U_1$ and $U_2$ which define its plaquette size along the two primitive lattice translation vectors $\vec{a}_1,\ \vec{a}_2$. The shape of a CLS plaquette is encoded by a $U_1\times U_2$ matrix $T$ with integer entries $0$ or $1$. The zero elements of the matrix $T$ prescribe the locations of unit cells with zero wavefunction amplitudes in the CLS plaquette. The number of all possible nontrivial matrices $T$ is finite and can be sorted and counted using an integer $s \geq 0$. Therefore we arrive at the extended CLS flatband classifier vector $\mus=(U_1, U_2, s)$.

For $U_1=1$ or $U_2=1$ there is only one possible shape and we can shorten the classifier $\mus$ to $(1,U_2)$ or $(U_1,1)$. The other cases considered below correspond to $U_1=U_2=2$, and we choose the integer $s$ to count the number of zeros in the above matrix $T$:
\begin{gather}
    s=0 \; : \;\;T=
    \begin{pmatrix}
        1 & 1 \\ 1 & 1
    \end{pmatrix},\\
    s=1 \; : \;\; T=
    \begin{pmatrix}
        1 & 0 \\ 1 & 1
    \end{pmatrix},\\
    s=2 \; : \;\; T=
    \begin{pmatrix}
        0 & 1 \\ 1 & 0
    \end{pmatrix}.
\end{gather}

\begin{figure}
    \centering
    \includegraphics[width=0.8\columnwidth]{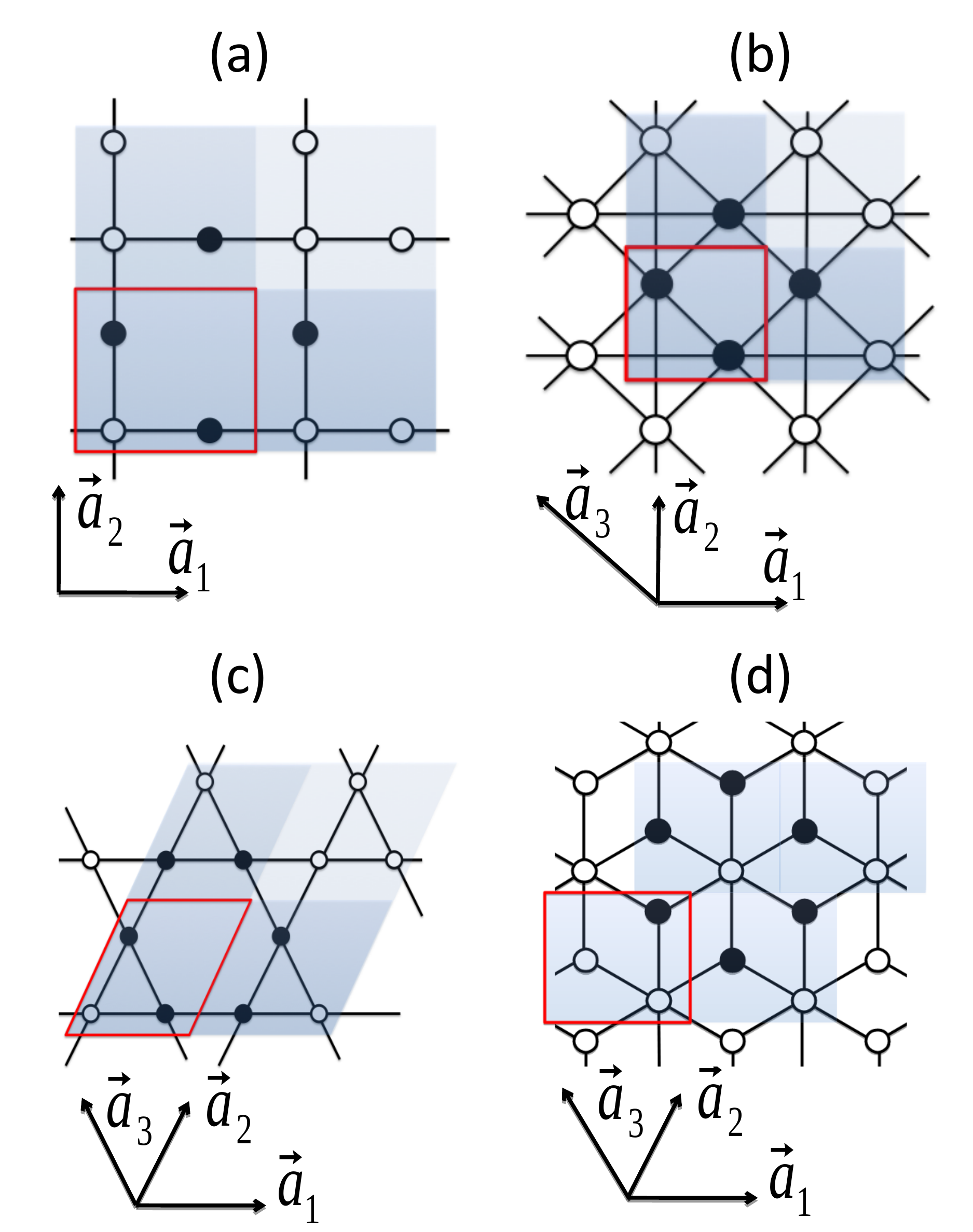}
    \caption{The $\mathbf{U}$-classification of $4$ specific and known 2D lattices with a flatband. Open circles denote the lattice sites, black lines nonzero hopping elements of equal value. Shaded areas indicate the $U_1\times U_2$ CLS plaquette, and the darker shaded regions show the occupied unit cells (black circles show the nonzero wavefunction amplitudes
of the flatband CLS). Red boxes denote the unit cell, and $\vec{a}_1,\ \vec{a}_2$ are primitive translation vectors, and $\vec{a}_{3}=\vec{a}_1 \pm \vec{a}_2$. Lattices and respective (extended) classifier vectors of the respective CLS:  (a) Lieb: $\mus=(2,2,1)$, (b) checkerboard: $\mus=(2,2,1)$, (c) kagome: $\mus=(2,2,1)$, and (d) dice: $\mus=(2,2,0)$.
}
    \label{fig:2d-known-eg-u-class}
\end{figure}

The case $U=1$ discussed in the introduction corresponds to $\mus=(1,1)$. The simplest nontrivial case in $d=2$ is $\mus=(2,1)$ and will be discussed below. The next and less trivial set of cases in $d=2$ is $\mus=(2,2,0)$, $\mus=(2,2,1)$ and $\mus=(2,2,2)$. Figure~\ref{fig:2d-known-eg-u-class} shows some of the known cases of 2D flatband networks with their CLS plaquettes: (a) Lieb: $\mus=(2,2,1)$, (b) checkerboard: $\mus=(2,2,1)$, (c) kagome: $\mus=(2,2,1)$, and (d) dice: $\mus=(2,2,0)$.

We will show that these known flatband networks are members of vast families of flatband networks, each with a number of continuously tunable control parameters. For $U_1,U_2 \leq 2$ the wavefunctions $\vpsi_n$ in Eq.~\eqref{eq:2d-nn-mat-rep} are nonzero for a maximum of four unit cells. We label these CLS components as $\vpsi_{i=1,\dots,4}$ as shown in Fig.~\ref{fig:u22-cls-config}(e). We will use the vector $\vpsi_i$ and the bra-ket $\ket{\psi_i}$ notations interchangeably throughout the text. 

\section{The flatband generator}

\subsection{The eigenvalue problem}

Just as in the $d=1$ case~\cite{maimaiti2019universal} we construct FB Hamiltonians from their CLS, considering the latter as  input parameters and reformulating the problem of finding a FB Hamiltonian into an inverse eigenvalue problem for the hopping matrices $H_\chi$. To achieve this we rewrite the eigenvalue problem~\eqref{eq:2d-nn-mat-rep} for the nonzero amplitudes $\vpsi_i$ and supplement it with destructive interference conditions which ensure the strict compactness of the eigenstate. Overall the way we solve this system in $d=2$ is very similar in spirit to but is in general more complex and involved than the $d=1$ case.

\subsubsection{$U=1$}

The $U=1$ case assumes a CLS which occupies only one unit cell with wave function $\vpsi_1$ and leads to a simple set of equations:
\begin{equation}
    \begin{aligned}
        H_0 \vpsi_1 & = \efb \vpsi_1, \\
        H_i \vpsi_1 & = H_i^\dagger \vpsi_1 = 0, \qquad i=1,2,3 \;.
    \end{aligned}
    \label{eq:u1-gen-eig-prob}
\end{equation} 

\subsubsection{$U_1=2$}

\begin{figure}
    \includegraphics[width=0.9\columnwidth]{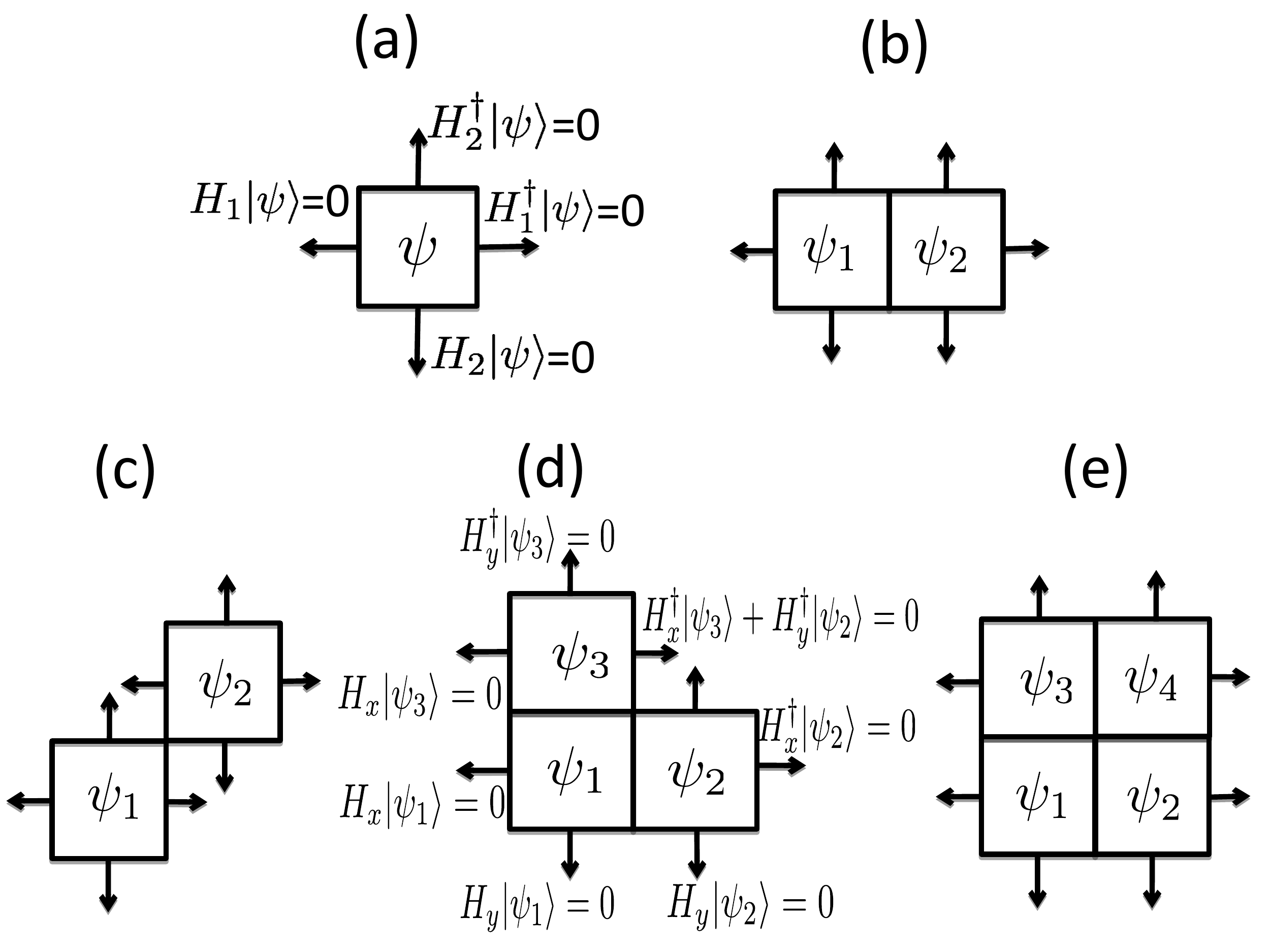}
    \caption{Classification of compact localised states for cases with two hopping matrices. Each square represents a unit cell. Directions of the hopping and respective destructive interference conditions are indicated by arrows. Where two hopping terms (arrows) meet, both will contribute to the destructive interference cancellation. (a) $\mus=(1,1)$ single unit cell  ($U=1$) CLS. (b) $\mus=(2,1)$ case. (c) $\mus=(2,2,2)$ case. (d) $\mus=(2,2,1)$ case. (e) $\mus=(2,2,0)$ case.}
    \label{fig:u22-cls-config} 
\end{figure}

\emph{Two hopping matrices} -- The simplest case of two hopping matrices $H_1,\ H_2$ can be always related to a square lattice geometry. One example is the Lieb lattice shown in Fig.~\ref{fig:2d-known-eg-u-class}(a). The possible CLS shapes and the hoppings for this case are shown in Fig.~\ref{fig:u22-cls-config}. The eigenvalue problem and destructive interference conditions read: 
\begin{equation}
    \begin{aligned}
        H_1 \vpsi_2 + H_2 \vpsi_3 \delta_{U_2,2} &= (\efb - H_0) \vpsi_1, \\
        H_1^\dagger \vpsi_1 + H_2 \vpsi_4 \delta_{U_2,2} \delta_{s,0} &= (\efb - H_0) \vpsi_2, \\ 
        \left(H_1 \vpsi_4 \delta_{s,0} + H_2^\dagger \vpsi_1 \right) \delta_{U_2,2}  &= (\efb - H_0)\vpsi_3\delta_{U_2,2}, \\ 
        \left(H_1^\dagger \vpsi_3 \delta_{s,0} + H_2^\dagger \vpsi_2 \delta_{s,0} \right) \delta_{U_2,2}  &= (\efb - H_0) \vpsi_4 \delta_{U_2,2} \delta_{s,0}, \\
        H_1 \vpsi_1 = H_1^\dagger \vpsi_2 &= 0, \\
        H_2 \vpsi_1 = H_2 \vpsi_2 &= 0, \\
        H_1 \vpsi_3 \delta_{U_2,2} = H_2^\dagger \vpsi_3 \delta_{U_2,2}  &= 0 , \\
        H_1 \vpsi_3 \delta_{U_2,2} \delta_{s,1} + H_2^\dagger \vpsi_2 &= 0 , \\
        H_1^\dagger \vpsi_4 \delta_{U_2,2} \delta_{s,0} = H_2^\dagger \vpsi_4 \delta_{U_2,2} \delta_{s,0} &= 0 \;.
    \end{aligned}
    \label{eq:u22-gen-eig-prob}
\end{equation} 
For specific values of $U_1,\ U_2$ and $s$ the above system gives the eigenvalue problem and destructive interference conditions for a CLS with the extended classifier vector $\mus=(U_1,U_2,s)$. 

\begin{figure}
    \includegraphics[width=0.9\columnwidth]{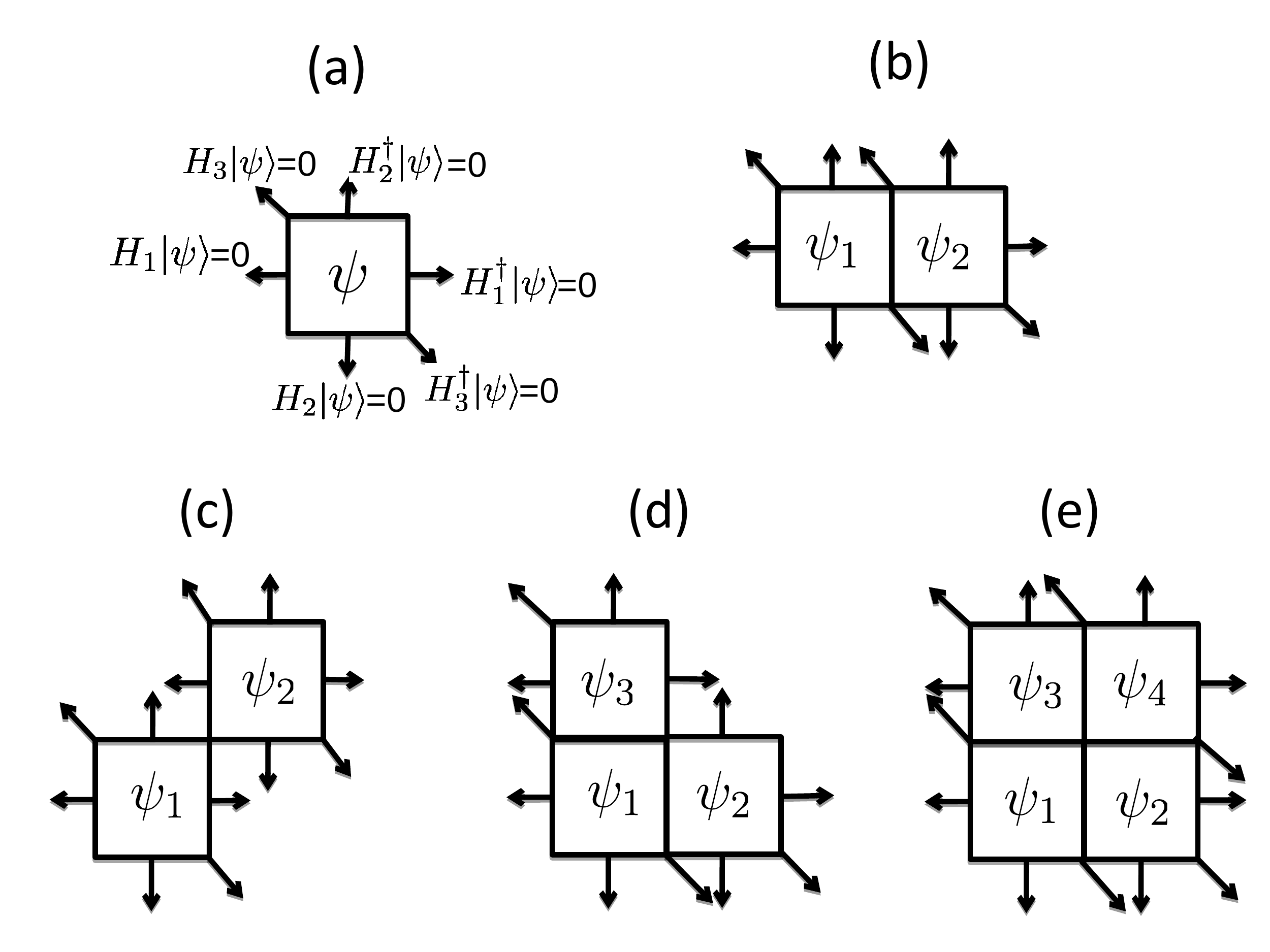}
    \caption{Classification of compact localised states for cases with three hopping matrices. Each square represents a unit cell. Directions of the hopping and respective destructive interference conditions are indicated by arrows. Where two or more hopping terms (arrows) meet, all will contribute to the destructive interference cancellation. (a) $\mus=(1,1)$ single unit cell ($U=1$) CLS. (b) $\mus=(2,1)$ case. (c) $\mus=(2,2,2)$ case. (d) $\mus=(2,2,1)$ case. (e) $\mus=(2,2,0)$ case.}
    \label{fig:u22-nnn-config}
\end{figure}

\emph{Three hopping matrices} -- Several known flatband lattice models have three hopping matrices which connect different unit cells, for example checkerboard, kagome, and dice lattices (see Fig.~\ref{fig:2d-known-eg-u-class}(b-d)). The three respective hoppings matrices $H_1,H_2,H_3$ as shown in Fig.~\ref{fig:u22-nnn-config}. It is instructive to note that the freedom in choosing different unit cells has consequences in our classification scheme. As an example, the dice lattice with its unit cell choice in Fig.~\ref{fig:2d-known-eg-u-class}(d) falls into the category $\mus=(2,2,0)$ with three nontrivial hopping matrices. The unit cell choice used in Fig.~1 in Ref.~\onlinecite{kolovsky2018topological} leads to a much larger CLS plaquette with classifier $\mus=(3,3)$ (and three empty unit cells in the CLS), but also to a reduced number of only two nontrivial hopping matrices.

For three hopping matrices the eigenvalue problem and the corresponding destructive interference conditions read:
{\footnotesize
\begin{equation}
    \begin{aligned}
        H_1 \vpsi_2 + H_2 \vpsi_3 \delta_{U_2,2} &= (\efb - H_0) \vpsi_1 , \\
        H_1^\dagger \vpsi_1 + H_2 \vpsi_4 \delta_{U_2,2} \delta_{s,0} + H_3^\dagger \vpsi_3 \delta_{U_2,2} &= (\efb - H_0) \vpsi_2, \\
        \left( H_1 \vpsi_4 \delta_{s,0} + H_2^\dagger \vpsi_1 + H_3 \vpsi_2 \right) \delta_{U_2,2} &= (\efb - H_0) \vpsi_3 \delta_{U_2,2}, \\
        \left( H_1^\dagger \vpsi_3 + H_2^\dagger \vpsi_2 \right) \delta_{U_2,2} \delta_{s,0} &=(\efb - H_0) \vpsi_4 \delta_{U_2,2} \delta_{s,0}, \\
        H_1 \vpsi_1 = H_2 \vpsi_1 &= 0, \\
        H_1^\dagger \vpsi_4 \delta_{U_2,2} \delta_{s,0} = H_2 \vpsi_4  \delta_{U_2,2} \delta_{s,0} &= 0,\\
        H_3^\dagger \vpsi_2 = H_3 \vpsi_3 \delta_{U_2,2} &= 0, \\
        H_2 \vpsi_2 + H_3^\dagger \vpsi_1 &= 0, \\
        H_1 \vpsi_3 \delta_{U_2,2} + H_3 \vpsi_1 \delta_{U_2,2} &= 0, \\
        H_2^\dagger \vpsi_3 \delta_{U_2,2} + H_3 \vpsi_4 \delta_{U_2,2} \delta_{s,0} &= 0, \\
        H_1^\dagger \vpsi_2 + H_3 \vpsi_4 \delta_{U_2,2} \delta_{s,0} &= 0 \; .
    \end{aligned}
    \label{eq:u22-nnn-eig-prob}
\end{equation} 
}
For specific values of $U_1,\ U_2,\ s$ we obtain the eigenvalue problem and destructive interference conditions for a CLS with the extended classifier vector $\mus=(U_1,U_2,s)$. In order to apply the generator defined below, we define the transverse projectors $Q_i, Q_{ij}, Q_{ijk}$ on $\{ \vpsi_i \}$, $\{ \vpsi_i,\vpsi_j\}$ and $\{ \vpsi_i, \vpsi_j, \vpsi_k \}$ respectively.

\subsection{The generator} 
\label{sec:2d-fb-gen}

The sets of equations~(\ref{eq:u22-gen-eig-prob},\ref{eq:u22-nnn-eig-prob}) are the starting point of our flatband generator. Our goal is to generate all possible matrices $H_{\chi}$ which allow for the existence of a flatband, given a particular choice of the CLS shape, $\efb$ and $H_0$. We arrive at the following protocol:
\begin{enumerate}
    \item Choose the number of bands $\nu$.
    \item Choose a hopping range.
    \item Choose a plaquette shape of the CLS. 
    \item Choose an arbitrary Hermitian $H_0$. 
    \item Choose a flatband energy $\efb$.
    \item Exclude $H_1,\ H_2$ and $H_3$ from the equations~(\ref{eq:u22-gen-eig-prob},\ref{eq:u22-nnn-eig-prob}) to get nonlinear constraints on the CLS components $\vpsi_i$, and solve these constraints to find all CLS components $\vpsi_i$.
    \item With the chosen $H_0, \efb$ and the CLS $\vpsi_i$ obtained at the previous step, solve equations~(\ref{eq:u22-gen-eig-prob},\ref{eq:u22-nnn-eig-prob}) for $H_{\chi}$.
\end{enumerate}
The above protocol admits variations which can simplify the task. In some cases the nonlinear constraints allow to skip item $5$ and keep the flatband energy $\efb$ a free parameter to be fixed when executing item $6$, as we will show below. Step $6$ requires solving nonlinear equations which may have either no CLS solutions, or a manifold of CLS solutions with freely tunable parameters, or even several such manifolds. Using a CLS solution from step $6$, and executing step $7$, will in general yield a solution manifold of hopping matrices with freely tunable parameters as well, which correspond to the freedom of fixing only the flatband states and not constraining the rest of the spectrum, bands, and eigenvectors.

\section{Solutions}
\label{sec:results}

We consider CLS sizes restricted to $U_1,U_2\leq 2$, unless stated otherwise.  

\subsection{$U=1$}

Without loss of generality we assume that $H_0$ is diagonal with first diagonal entry $\efb$. From the first line in Eq. \eqref{eq:u1-gen-eig-prob} we conclude $\vpsi_1=(1,0,0,...,0)$. The following destructive interference conditions yield $(H_i)_{1,\mu}=(H_i)_{\mu,1}=0$, i.e. the hopping matrices have zeros on their first row and column, and freely choosable entries elsewhere. The entries parametrize the remaining dispersive degrees of freedom. These entries as well as an overall multiplicative scaling factor and energy gauge are the free parameters of the corresponding manifold of $U=1$ flatband Hamiltonians. These solutions correspond to the detangled basis of $U=1$ flatband networks.~\cite{flach2014detangling} Additional manifold parameters are obtained from entangling the CLS with the dispersive network through commuting local unitary operations.~\cite{flach2014detangling} 

For two bands $\nu=2$ and $H_3=0$ it follows that the only solutions are either $U=1$ flatbands, or decoupled 1D networks 
with $H_2=0$ and $U=2$ (see details of the derivation in Appendix~\ref{app:two-band-prob-2d}).

\subsection{$\mus=(2,1)$}

\begin{figure}
    \includegraphics[width=0.45\columnwidth]{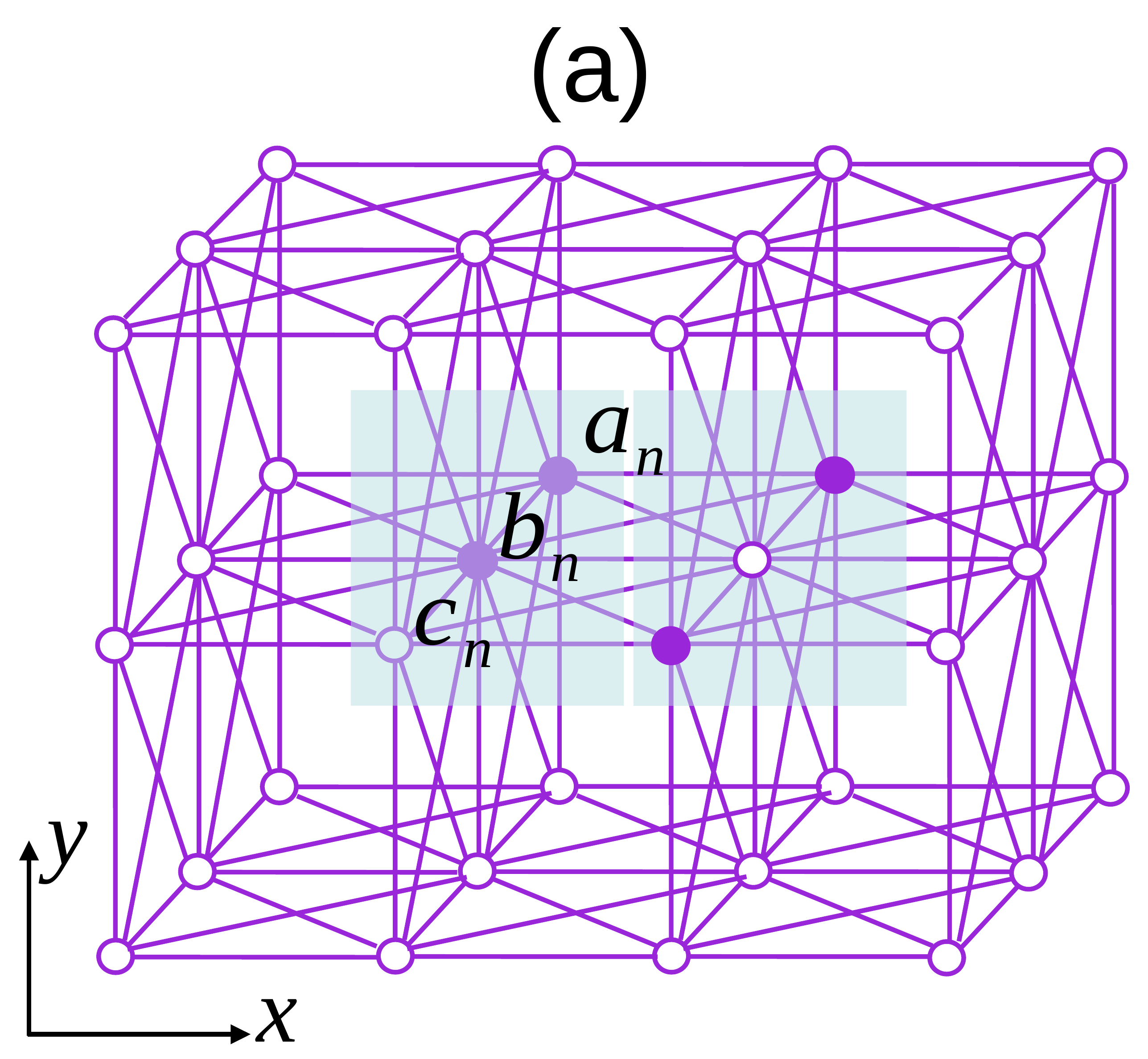}
    \includegraphics[width=0.5\columnwidth]{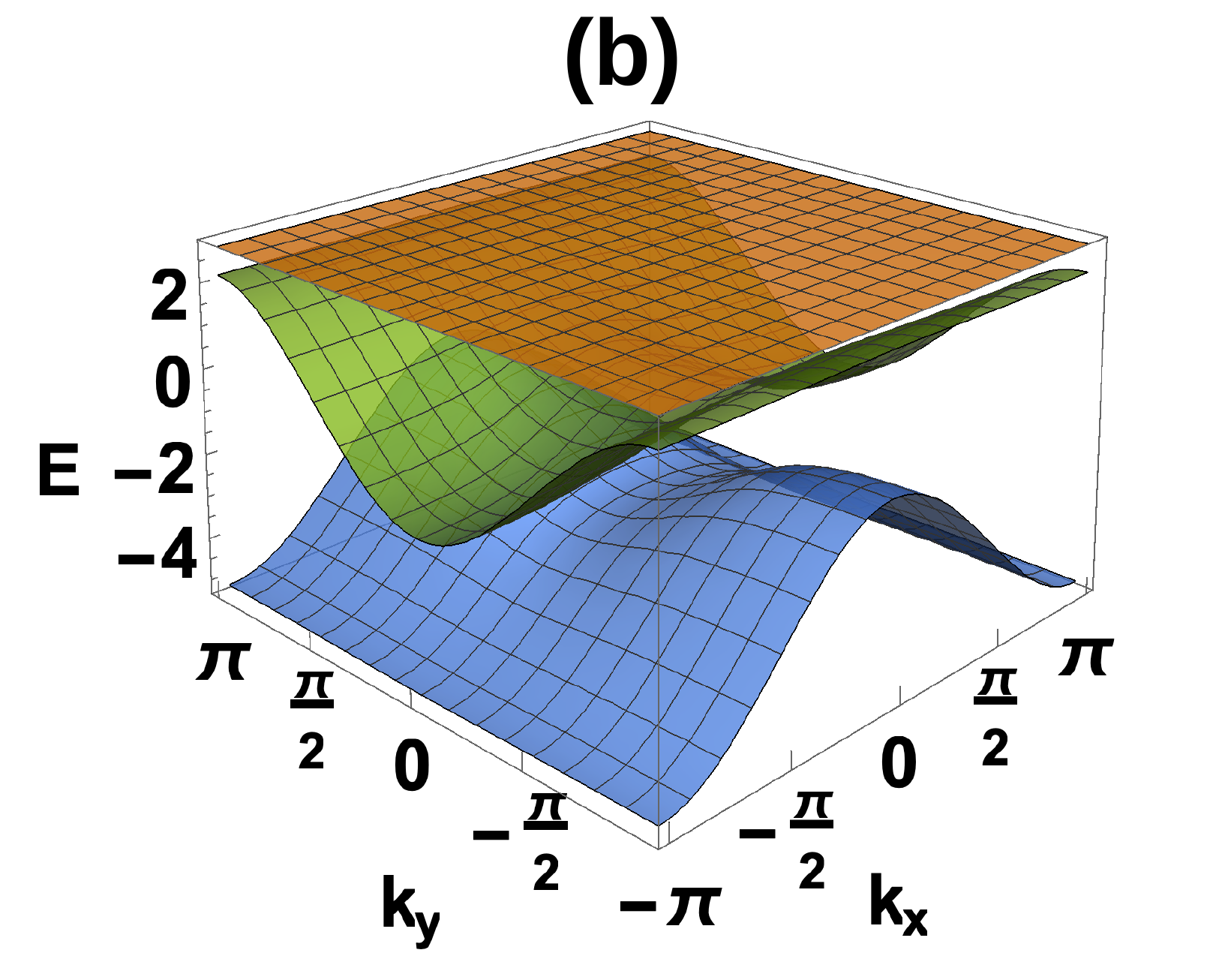}
    \caption{Example of a $\mus=(2,1)$ CLS FB network and $\nu=3$. (a) Tight binding lattice with two hopping matrices. Lines indicate nonzero hoppings, filled circles show the position of a CLS. $a_n,b_n,c_n$ indicate the sites in one unit cell. See Appendix ~\ref{app:nn-u21} for details. (b) Band structure corresponding to (a).}
    \label{fig:nn-u21-example}
\end{figure}

A schematic of the CLS and the destructive interference conditions for this case are shown in Fig.~\ref{fig:u22-cls-config}(b) and~Fig.~\ref{fig:u22-nnn-config}(b) for the two and three hopping matrices cases respectively. The eigenvalue problem~\eqref{eq:u22-gen-eig-prob}, respectively~\eqref{eq:u22-nnn-eig-prob}, involves only one hopping matrix $H_1$, while the matrices $H_{2,3}$ enter additional destructive interference conditions only. Therefore the eigenvalue problem reduces to the 1D case solved in Ref.~\onlinecite{maimaiti2019universal}. It follows (see Appendix~\ref{app:nn-u21} and Ref.~\onlinecite{maimaiti2019universal} for details)
\begin{gather}
    H_1 = \frac{(\efb - H_0)\ket{\psi_1}\bra{\psi_2}(\efb - H_0)}{\mel{\psi_1}{\efb - H_0}{\psi_1}}\;.
\end{gather}
The CLS components $\vpsi_1,\vpsi_2$ are subject to nonlinear constraints
\begin{equation}
    \begin{aligned}
        \mel{\psi_2}{\efb - H_0}{\psi_1} &= 0, \\
        \mel{\psi_1}{\efb - H_0}{\psi_1} &= \mel{\psi_2}{\efb - H_0}{\psi_2} \;,
    \end{aligned}
    \label{eq:u21-cls-cons}
\end{equation}
that are resolved similarly to the 1D case.~\cite{maimaiti2019universal} For two hopping matrices the additional destructive interference yields
\begin{gather}
    H_2 = Q_{12} M_2 Q_{12},
    \label{eq:u210-sol}
\end{gather}
where $M_2$ is an arbitrary $\nu \times \nu$ matrix. Figure~\ref{fig:nn-u21-example} provides an example network for the two hopping matrices case. For three hopping matrices the derivation of $H_2$ and $H_3$ is given in Appendix~\ref{app:nnn-u21}.

\subsection{$\mus=(2,2,2)$}

The case $\mus=(2,2,2)$ is shown in Figs.~\ref{fig:u22-cls-config}(c) and~\ref{fig:u22-nnn-config}(c), where the two CLS-occupied unit cells just touch one another. Equations~(\ref{eq:u22-gen-eig-prob}-\ref{eq:u22-nnn-eig-prob}) reduce to simple eigenproblems for the amplitudes in each individual unit cell:
\begin{gather}
    H_0 \kpsi{i=1,2} = \efb\kpsi{i=1,2}.
\end{gather} 
We choose some $H_0$ and thereby fix $\efb$ and $\vpsi_{1,2}$. If $\efb$ is nondegenerate, then $\vpsi_2\propto\vpsi_1$ and the problem reduces to a $U=1$ CLS. If $\efb$ is degenerate we need at least $\nu=3$ bands. The amplitudes $\vpsi_{1,2}$ can then be picked as distinct linear combinations of the eigenvectors corresponding to $\efb$. The intra-cell hopping matrices $H_1, H_2, H_3$ are reconstructed from the destructive interference conditions, Eqs.~(\ref{eq:u22-gen-eig-prob}-\ref{eq:u22-nnn-eig-prob}):
\begin{gather*}
    H_1\kpsi{1} = H_2\kpsi{1} = 0 \qquad \bpsi{2} H_1 = \bpsi{2} H_2 = 0 \\
    H_1^\dagger\kpsi{1} + H_2\kpsi{2} = 0 \qquad H_2^\dagger\kpsi{1} + H_1\kpsi{2} = 0 \\
    H_3\kpsi{1} = H_3^\dagger\kpsi{1} = H_3\kpsi{2} = H_3^\dagger\kpsi{2} = 0
\end{gather*}
The last line implies that  $H_3=Q_{12}MQ_{12}$ where $M$ is an arbitrary $\nu\times\nu$ matrix. The first two lines of the above equation constitute a coupled inverse problem of finding $H_1, H_2$ from their known action on $\vpsi_1,\vpsi_2$. This problem can be decoupled into inverse problems for $H_1$ and $H_2$ by defining:
\begin{gather}
    H_1\kpsi{1} = Q_1\ket{z}\qquad H_1\kpsi{2} = Q_2 \ket{w}.
\end{gather}
The inverse problems for $H_1, H_2$ have been solved in Ref.~\onlinecite{maimaiti2019universal}.

\subsection{$\mus=(2,2,1)$} 

A schematic of the CLS and destructive interference conditions for this case are shown in Fig.~\ref{fig:u22-cls-config}(d) and Fig.~\ref{fig:u22-nnn-config}(d) for the two and three hopping matrices respectively. 

\subsubsection{Two hopping matrices}

The case of two hopping matrices and an arbitrary number of bands can be resolved following a similar derivation as for $\mus=(2,1)$, however the solution is cumbersome. Therefore for simplicity we focus on the specific case of $\nu=3$ bands. We can consider two cases: (a) the CLS amplitudes are linearly independent, or (b) the CLS amplitudes are dependent. The latter case includes the known cases of the Lieb~\cite{lieb1989two} and  Tasaki~\cite{tasaki1992ferromagnetism,mielke1993ferromagnetism} lattices.

Case (a) has one of the destructive interference conditions reading $\bpsi{3} H_1 + \bpsi{2}H_2 = 0$.  For $\nu=3$ the number of components of each of the vectors $\Psi_{1,2,3}$ is also equal to three. Therefore it is straightforward to show that the destructive interference condition splits into two:
\begin{gather}
    \begin{aligned}
  	    \bpsi{3} H_1 = & \bpsi{2} H_2 = 0 .
    \end{aligned}
\end{gather}
Then the eigenvalue problem Eq.~\eqref{eq:u22-gen-eig-prob} reads
\begin{equation}
    \begin{aligned}
        & H_1 \vpsi_2 + H_2 \vpsi_3 = (\efb -H_0) \vpsi_1, \\
        & H_1^{\dagger} \vpsi_2 = (\efb - H_0) \vpsi_2, \\
        & H_2^{\dagger} \vpsi_2 = (\efb - H_0) \vpsi_3, \\
        & H_1 \vpsi_2 = H_1^{\dagger} \vpsi_2 = H_1 \vpsi_3 = H_1^{\dagger} \vpsi_3 = 0, \\
        & H_2 \vpsi_2 = H_2^{\dagger} \vpsi_3 = H_2 \vpsi_2 = H_2^{\dagger} \vpsi_2 = 0.
    \end{aligned}  
    \label{eq:lshap-eig-prob}
\end{equation} 
We eliminate $H_1,H_2$ from the eigenproblem and obtain the nonlinear constraints on the CLS amplitudes:
\begin{equation}
    \begin{aligned}
        & \mel{\psi_2}{H_0}{\psi_1} = \efb \bra{\psi_2}\ket{\psi_1}, \\
        & \mel{\psi_3}{H_0}{\psi_1} = \efb \bra{\psi_3}\ket{\psi_1}, \\
        & \mel{\psi_3}{H_0}{\psi_2} = \efb \bra{\psi_3}\ket{\psi_2}, \\
        & \mel{\psi_2}{\efb - H_0}{\psi_2} + \mel{\psi_3}{\efb - H_0}{\psi_3} \\
        &= \mel{\psi_1}{\efb - H_0}{\psi_1}.
    \end{aligned}
    \label{eq:non-lin-const-L-shape-app}
\end{equation} 
Finally we obtain the hopping matrices:
\begin{equation}
    \begin{aligned}
        H_1 & = \frac{(\efb - H_0) \ket{\psi_1} \bra{\psi_2} (\efb - H_0)}{\mel{\psi_1}{\efb - H_0}{\psi_1}}, \\
        H_2 & = \frac{(\efb - H_0) \ket{\psi_1} \bra{\psi_3} (\efb - H_0)}{\mel{\psi_1}{\efb - H_0}{\psi_1}}.
    \end{aligned} 
    \label{eq:u221-spe-sol}
\end{equation} 

Case (b) assumes $\vpsi_{i=1,2,3}$ to be linearly dependent
\begin{equation}
    \label{eq:nu3-lin-dep-cls-comp-1}    
    \ket{\psi_1} = \alpha \ket{\psi_2} + \beta \ket{\psi_3}.
\end{equation}
This yields the following solution (see details in Appendix~\ref{app:nn-u221}): 
\begin{equation}
    \begin{aligned}
        H_1 &= \frac{Q_2\ket{a}\bra{\psi_2}(\efb - H_0)}{\mel{\psi_3}{Q_2}{a}}, \\
        H_2 &= \frac{Q_3\ket{b}\bra{\psi_3}(\efb - H_0)}{\mel{\psi_2}{Q_3}{b}},
    \end{aligned}
    \label{eq:u221-hxy-sol-3band-1}
\end{equation} 
where $\ket{a}, \ket{b}$ are arbitrary vectors, and $\efb, H_0, \vpsi_2, \vpsi_3$ are chosen respecting the constraints \begin{equation}
    \begin{aligned}
        \mel{\psi_2}{\efb - H_0}{\psi_2} &= 0, \\
        \mel{\psi_3}{\efb - H_0}{\psi_3} &= 0, \\
        \mel{\psi_3}{\efb - H_0}{\psi_2} &= 0, \\
        \left(\efb - H_0\right)\left(\alpha\ket{\psi_2} + \beta\ket{\psi_3}\right) &= 0.
    \end{aligned}
    \label{eq:u221-cls-cons-3band-1}
\end{equation} 
Equations~(\ref{eq:u221-hxy-sol-3band-1}, \ref{eq:u221-cls-cons-3band-1}, \ref{eq:nu3-lin-dep-cls-comp-1}) provide a complete solution to this special $\mus=(2,2,1)$ case with $\nu=3$ bands. The known examples such as Lieb lattice and Tasaki's lattice can be constructed from our generator as demonstrated in Appendix~\ref{app:nn-u221}. Figure~\ref{fig:nn-u221-example}(a,b) shows one generated example (see details in Appendix~\ref{app:nn-u221}).

\begin{figure}
	\includegraphics[width=0.5\columnwidth]{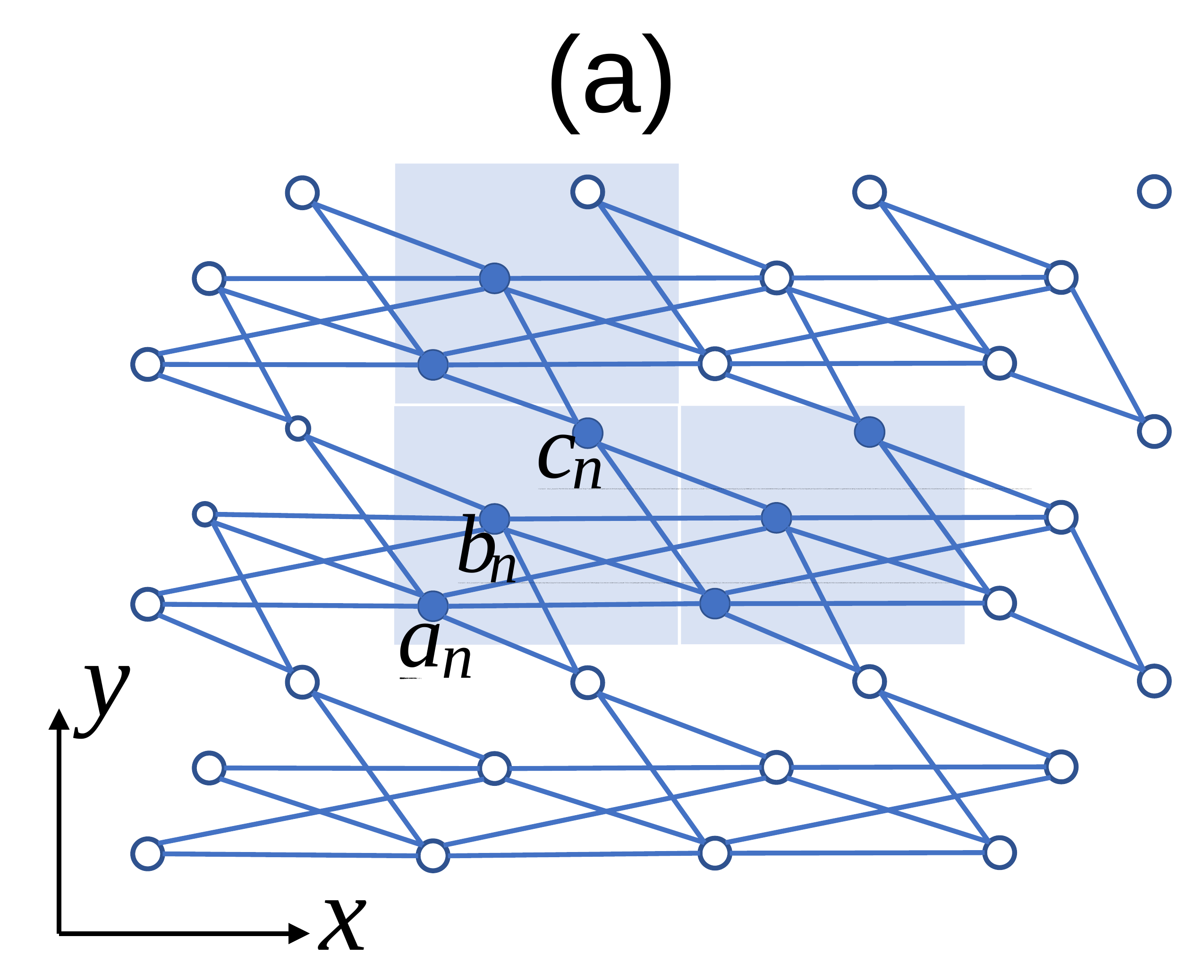}
	\includegraphics[width=0.45\columnwidth]{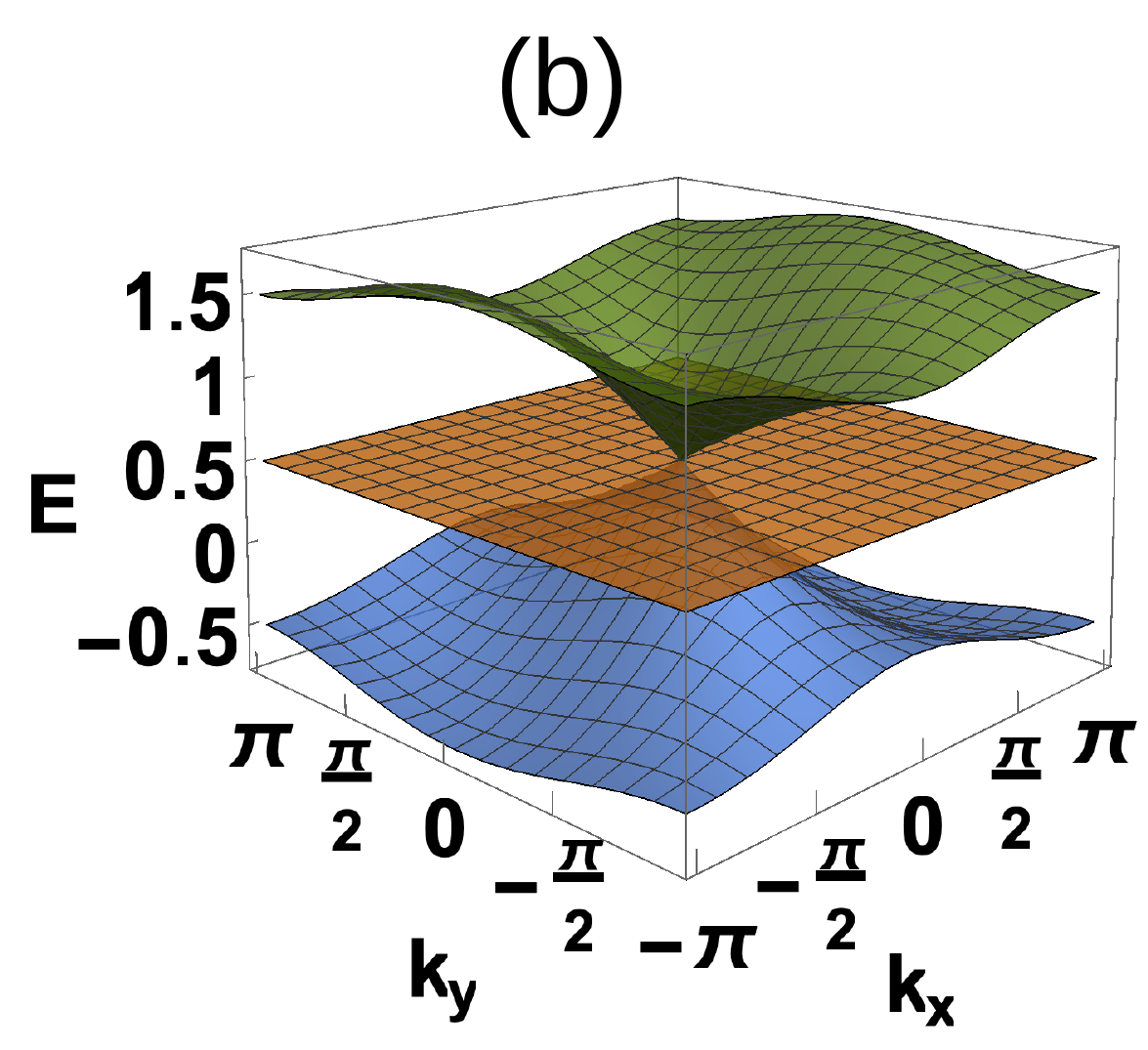}
	\includegraphics[width=0.45\columnwidth]{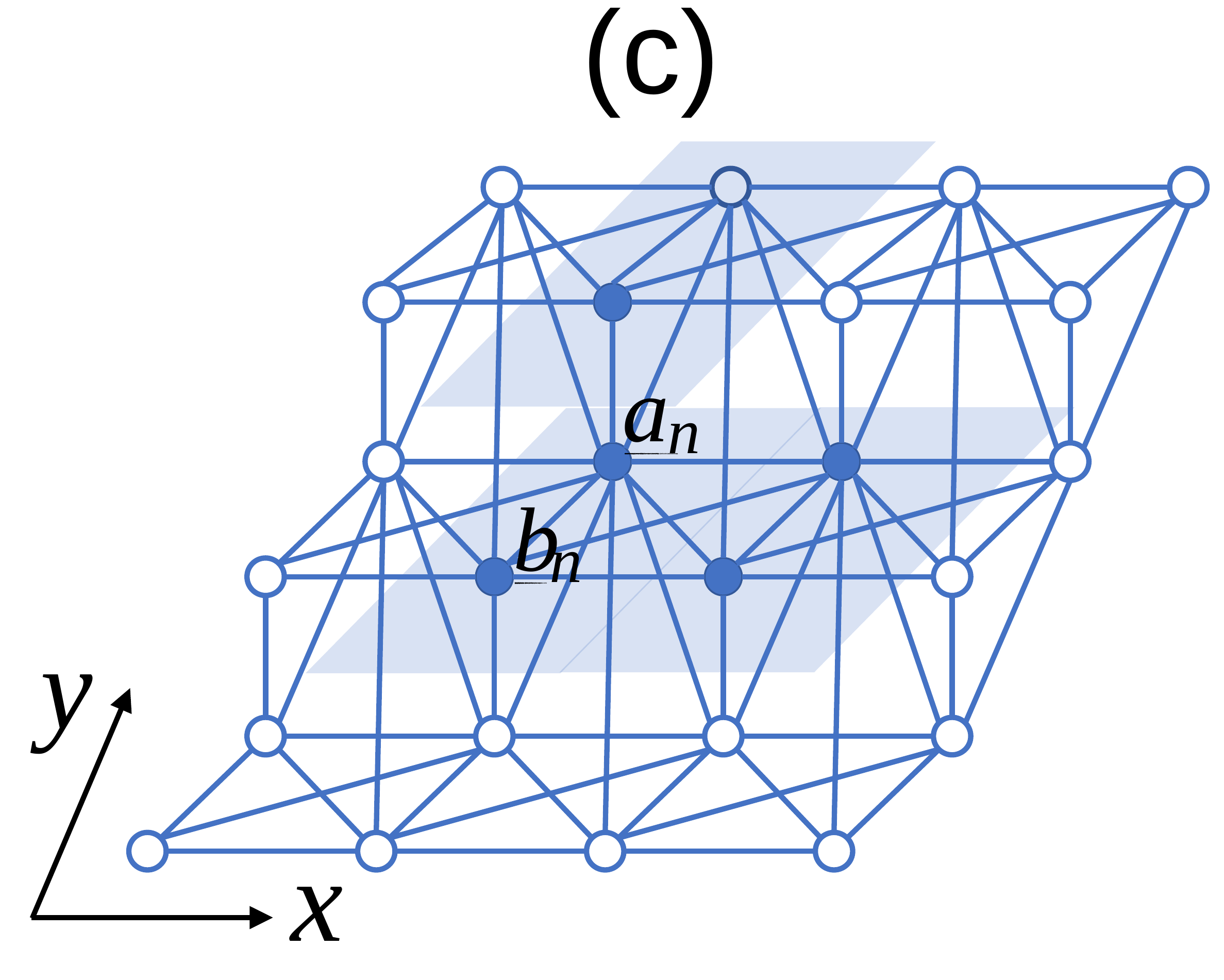}
	\includegraphics[width=0.48\columnwidth]{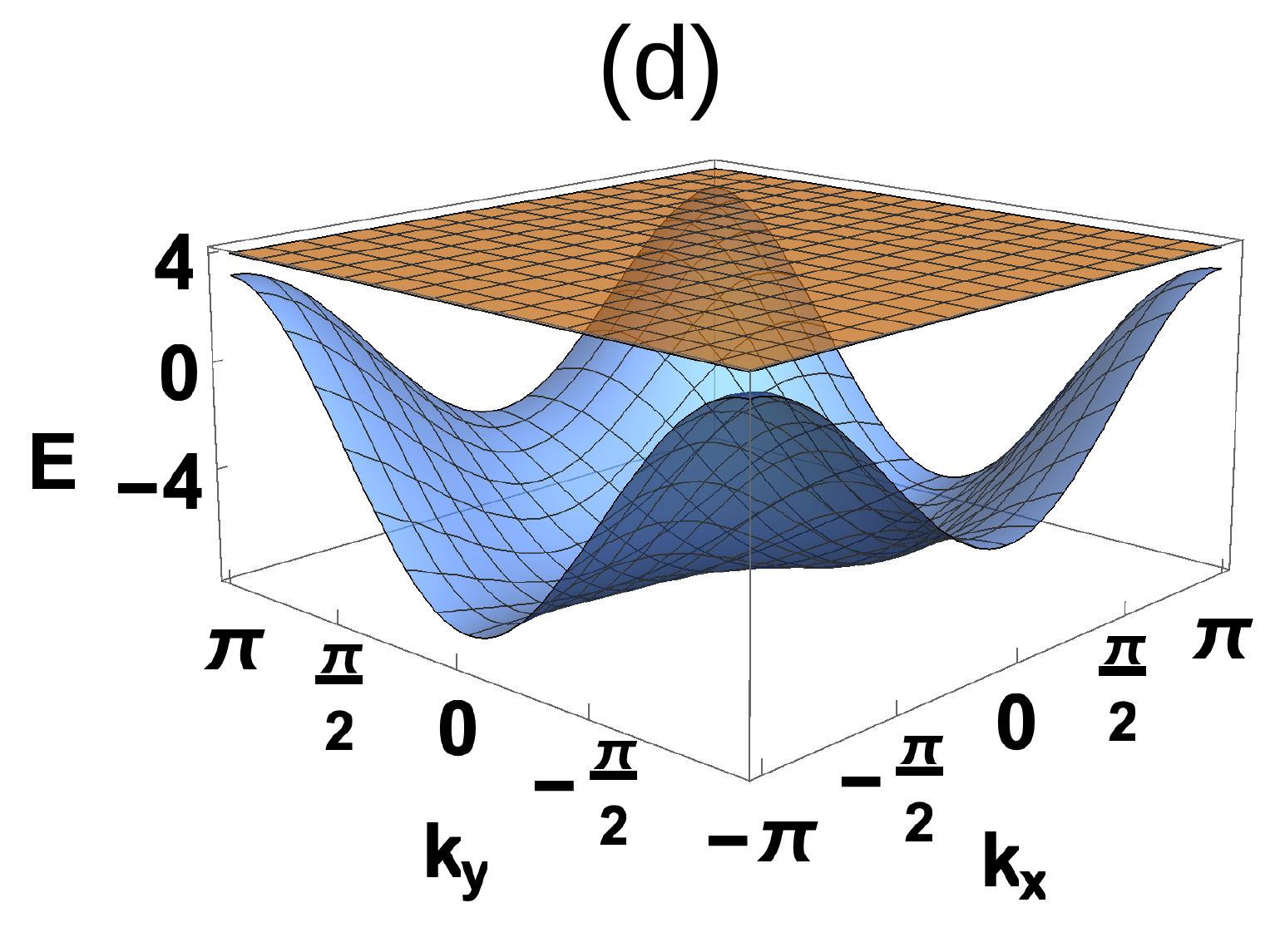}
	\caption{Examples of $\mus=(2,2,1)$ CLS FB networks. (a) Tight binding lattice with two hopping matrices and $\nu=3$. Lines indicate nonzero hoppings, filled circles show the position of a CLS. $a_n,b_n,c_n$ indicate the sites in one unit cell. See Appendix ~\ref{app:nn-u221} for details. (b) Band structure corresponding to (a). (c) Same as (a) but for three hopping matrices and $\nu=2$. $a_n,b_n$ indicate the sites in one unit cell. See Appendix ~\ref{app:nnn-u221} for details. (d) Band structure corresponding to (c).}
	\label{fig:nn-u221-example}
\end{figure}

\subsubsection{Three hopping matrices}

The configuration of this case is shown in Fig.~\ref{fig:u22-nnn-config}(d). The eigenvalue problem and the destructive interference conditions read
\begin{equation}
    \begin{aligned}
       H_1 \kpsi{2} + H_2 \kpsi{3} &= (\efb - H_0) \kpsi{1}, \\
       H_1^\dagger \kpsi{1} + H_3^\dagger \kpsi{3} &= (\efb - H_0) \kpsi{2} , \\
       H_2^\dagger \kpsi{1} + H_3 \kpsi{2} &= (\efb - H_0) \kpsi{3} ,\\
       H_1 \kpsi{1} &= H_2 \kpsi{1} = H_3 \kpsi{3} = 0,\\
       \bpsi{2} H_1 &= \bpsi{3} H_2 = \bpsi{2} H_3 = 0, \\
       H_1 \kpsi{3} + H_3 \kpsi{1} &= 0, \\
       H_2 \kpsi{2} + H_3^\dagger \kpsi{1} &= 0, \\
       \bpsi{3} H_1 + \bpsi{2} H_2 &= 0 \; .
    \end{aligned}
    \label{eq:nnn-triang-eig-prob}
\end{equation} 
The family of flatband solutions for two bands $\nu=2$ is derived in Appendix~\ref{app:nnn-u221}, with a particular member choice of the family shown in Fig.~\ref{fig:nn-u221-example}(c,d). Notably the checkerboard lattice~\cite{mielke1991ferromagnetic,mielke1991ferromagnetism,palmer2001quantum,canals2002from} is also part of the family, as outlined in Appendix~\ref{app:nnn-u221}. For $\nu=3$ a more cumbersome derivation yields the family of flatband solutions which contains the kagome lattice ~\cite{mamoru2003kagome,itiro1951statistics,mielke1992exact} (not shown here).

\subsection{$\mus=(2,2,0)$} 

\begin{figure}
	\includegraphics[width=0.45\columnwidth]{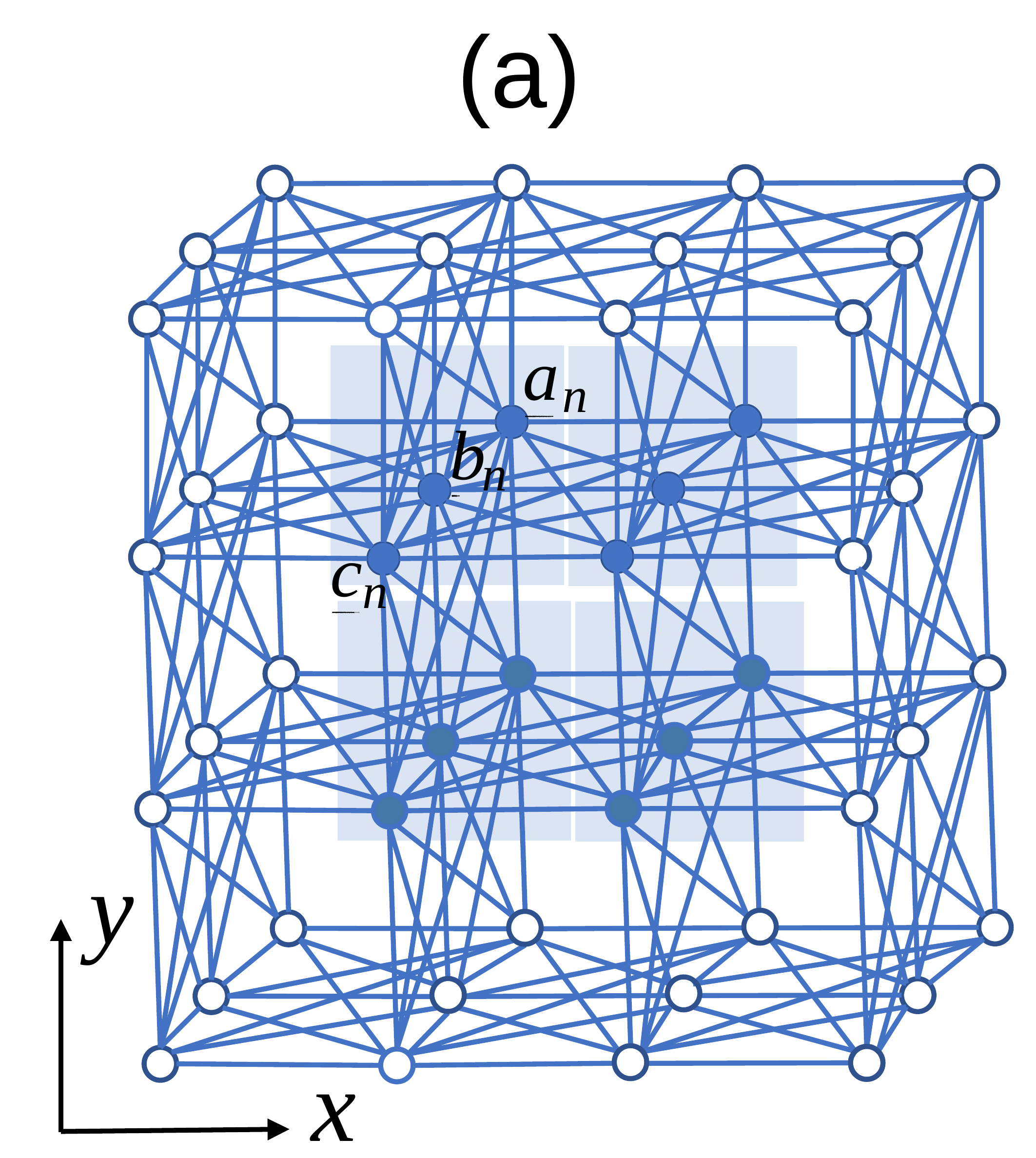}
	\includegraphics[width=0.45\columnwidth]{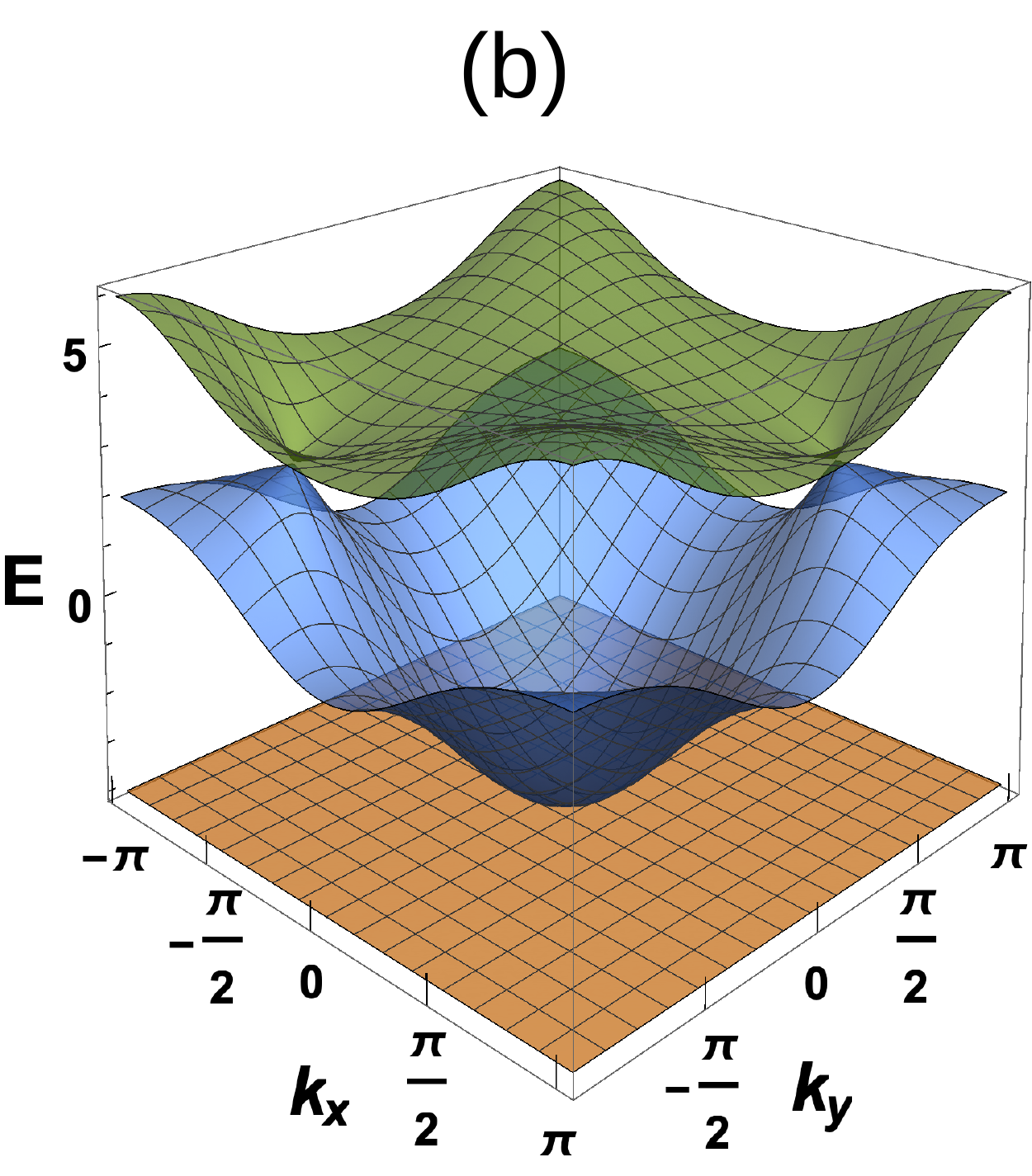}
	\caption{Example of a $\mus=(2,2,0)$ CLS FB network, two hopping matrices, and $\nu=3$. (a) Tight binding lattice. Lines indicate nonzero hoppings, filled circles show the position of a CLS. $a_n,b_n,c_n$ indicate the sites in one unit cell. See Appendix ~\ref{app:nn-u220-nu3} for details. (b) Band structure corresponding to (a).}
	\label{fig:nn-u220-example}
\end{figure}

The CLS and destructive interference conditions for the case of two hopping matrices and the square shaped CLS are shown in Fig.~\ref{fig:u22-cls-config}(e). For simplicity we restrict ourselves to three bands and use a direct parameterization of the CLS amplitudes $\vpsi_i$ to solve Eq.~\eqref{eq:u22-gen-eig-prob}. The full analytical solution is reported in Appendix~\ref{app:nn-u220-nu3}. There are three free parameters in this solution, and Fig.~\ref{fig:nn-u220-example} shows an example FB Hamiltonian for this case. 

To conclude we note that the increased complexity of the equations~(\ref{eq:u22-gen-eig-prob}-\ref{eq:u22-nnn-eig-prob}) for fully 2D shapes as compared to the 1D case is an expected and generic feature. However while it does not seem to be possible to work out full solutions in general -- the nonlinear constraints on the amplitudes and complex destructive interference conditions typically being the main obstacle -- we believe that such solutions can be found in individual cases.

\section{Conclusions} 
\label{sec:conclusion}

In this work, we extended the systematic 1D flatband generator~\cite{maimaiti2019universal} to two dimensions. 
Two important additional classifiers have been identified and added to make the 2D generator complete. First, we need to specify the
underlying Bravais lattice class. Second and most importantly we need to specify the shape of the compact localized states at otherwise fixed
CLS plaquette size. We derived analytical solutions for a number of different FB classes, and reproduced some of the well-known FB lattices: Lieb, Tasaki, kagome and checkerboard, along with a number of new 2D FB lattice examples. Our generator results in the possibility of counting free continuously variable parameters  after fixing one particular 2D FB class. These existing parameters demonstrate that FB Hamiltonians, while being fine-tuned models, emerge as members of finite dimensional Hamiltonian manifolds with an additional rich structure. Our results can be straightforwardly extended to larger compact localized states in 2D, and also to 3D cases, no matter how cumbersome the derivations could turn. Therefore our FB generator provides a direct way to search for flatbands for fixed lattice geometries in any lattice dimension.

\begin{acknowledgments}
    This work was supported by the Institute for Basic Science in Korea (IBS-R024-D1).
\end{acknowledgments}

\appendix

\section{$\nu=2$ and $H_3=0$}
\label{app:two-band-prob-2d} 

We consider two bands and $H_3=0$ and demonstrate that any possible flatbands always either reduce to class $U=1$ or decouple into 1D networks.

First we consider the $\mus=(2,1)$ case in Fig.~\ref{fig:u22-cls-config}(b). The eigenvalue problem and the destructive interference conditions~\eqref{eq:u22-gen-eig-prob} read
\begin{equation}
    \begin{aligned}
        & H_0 \vpsi_1 + H_1 \vpsi_2 = \efb \vpsi_1,\\
        & H_0 \vpsi_2 + H_1^{\dagger} \vpsi_1  = \efb \vpsi_2, \\
        & H_1 \vpsi_1 = H_1^{\dagger} \vpsi_2 = 0,  \\
        & H_2 \vpsi_{s} = H_2^{\dagger} \vpsi_{s} = 0, \ \  s=1, 2.
    \end{aligned}
    \label{ev_prob_nu2_u21}
\end{equation}
The last line enforces that either (i) $H_2=0$ or  (ii) $\vpsi_2 \propto \vpsi_1$. Case (i) reduces the system to a set of disconnected 1D networks which were completely studied in Ref.~\onlinecite{maimaiti2017compact}. Case (ii) yields that $\vpsi_{1,2}$ are an eigenvector to $H_0$ (up to a normalization  factor) and form a complete $U=1$ CLS, and the considered $\mus=(2,1)$ case reduces to a linear combination of two $U=1$ CLS states.

For the $\mus=(2,2,0)$ case in Fig.~\ref{fig:u22-cls-config}(e) the destructive interference conditions in Eq.~\eqref{eq:u22-gen-eig-prob} read
\begin{equation}
    \begin{aligned}
        & H_1 \vpsi_{1} = H_1 \vpsi_{3} = 0, \\
        & H_1^{\dagger} \vpsi_{2} =  H_1^{\dagger} \vpsi_{4} = 0,\\
        & H_2 \vpsi_{1} = H_2 \vpsi_{2} = 0,\\
        & H_2^{\dagger} \vpsi_{3} = H_2^{\dagger} \vpsi_{4} = 0.
    \end{aligned}
    \label{ev_prob_nu2_u22}
\end{equation}
The first line enforces that either (i) $H_1=0$ or  (ii) $\vpsi_3 \propto \vpsi_1$. Case (i) reduces the system to disconnected 1D networks. The third line results in (iia) $H_2=0$ or (iib) $\vpsi_2 \propto \vpsi_1$. Case (iia) reduces the system to disconnected 1D networks. Case (iib) implies $\vpsi_4 \propto \vpsi_1$ and reduces the problem to $U=1$.

The case $\mus=(2,2,1$) shown in Fig.~\ref{fig:u22-cls-config}(d) yields the following destructive interference conditions:
\begin{equation}
    \begin{aligned}
        & H_1 \vpsi_{1} = H_1 \vpsi_{3} = 0, \\
        & H_1^{\dagger} \vpsi_{2} = 0,\\
        & H_2 \vpsi_{1} = H_2 \vpsi_{2} = 0,\\
        & H_2^{\dagger} \vpsi_{3}  = 0 , \\
        & H_1^{\dagger} \vpsi_3 +H_2^{\dagger} \vpsi_2=0.
    \end{aligned}
    \label{ev_prob_nu2_u221}
\end{equation}
The first line enforces that either (i) $H_1=0$ or (ii) $\vpsi_3 \propto \vpsi_1$. Case (i) reduces the system to disconnected 1D networks. The third line results in (iia) $H_2=0$ or (iib) $\vpsi_2 \propto \vpsi_1$. Case (iia) reduces the system to disconnected 1D networks. Case (iib) reduces the problem to $U=1$.

The case $\mus=(2,2,2$) shown in Fig.~\ref{fig:u22-cls-config}(c) is slightly more involved. The eigenproblem in this case reads:
\begin{gather*}
    H_0\vpsi_{1,2} = \efb\vpsi_{1,2}.
\end{gather*}
There are two possible solutions: (i) $\psi_2\propto\psi_1$ or (ii) $H_0 = \efb\mathbb{I}$ and $\vpsi_2\perp\vpsi_1$. Case (i) reduces the system to $U=1$. In case (ii) we consider the destructive interference conditions
\begin{gather*}
    H_1\vpsi_1 = H_2\vpsi_1 = 0\\
    H_1^\dagger\vpsi_2 = H_2^\dagger\vpsi_2 = 0\\
    H_1\vpsi_2 + H_2^\dagger\vpsi_1 = 0\\
    H_2\vpsi_2 + H_1^\dagger\vpsi_1 = 0
\end{gather*}
From the first two lines and orthogonality of $\vpsi_{1,2}$ we conclude that $H_{1,2} \propto \ket{\psi_1}\bra{\psi_2}$. However this is incompatible with the remaining two destructive interference conditions as verified by direct substituion  and taking into account the mutual orthogonality of $\vpsi_{1,2}$.

\section{FB generation for two hopping matrices and $\nu \geq 3$}

\subsection{$U=(2,1)$} 
\label{app:nn-u21}

From Eq.~\eqref{eq:u22-gen-eig-prob}, we get the eigenvalue problem and destructive interference conditions 
\begin{equation}
    \begin{aligned}
       H_1 \kpsi{2} & = (\efb - H_0) \kpsi{1}, \\ 
       \bpsi{1} H_1 &= \bpsi{1} (\efb - H_0), \\ 
       H_1 \kpsi{1} &= H_2 \kpsi{1} = H_2 \kpsi{2} = 0, \\
       \bpsi{2} H_1 &= \bpsi{1} H_2 = \bpsi{2} H_2 = 0. \
    \end{aligned}
    \label{eq:u21-eig-prob-app}
\end{equation} 
Since $H_2$ only appears in the destructive interference conditions, we can express it as $H_2 = Q_{12} M_2 Q_{12}$ where $M_2$ is an arbitrary $\nu\times\nu$ matrix. The remaining problem is identical to the 1D problem discussed in our previous work~\cite{maimaiti2019universal} so that we only sketch the solution. Using the destructing interference conditions, we eliminate $H_1$ from the eigenvalue problem and find the following nonlinear constraints on the CLS
\begin{equation}
    \begin{aligned}
        \mel{\psi_2}{\efb - H_0}{\psi_1} &= 0, \\
        \mel{\psi_1}{\efb - H_0}{\psi_1} &= \mel{\psi_2}{\efb - H_0}{\psi_2}. \
    \end{aligned}
    \label{eq:u21-cls-cons-app}
\end{equation} 
The destructive interference conditions suggest the following ansatz $H_1 = Q_2 \ket{u} \bra{v} Q_1$, where $\vert u \rangle, \vert v \rangle$ are vectors to be fixed. Plugging this ansatz into Eq.~\eqref{eq:u21-eig-prob-app} we find the vectors $\vec{u},\vec{v}$ and the final expression for $H_1$
\begin{gather}
    \label{eq:u21-hx-hy-def-app-final}
    H_1 = \frac{(\efb - H_0)\ket{\psi_1} \bra{\psi_2}(\efb - H_0)}{\mel{ \psi_1}{\efb - H_0}{\psi_1}}.
\end{gather}

\subsubsection{$\nu=3$ example}
\label{app:nn-u21-eg}

We choose $H_0$ and we parameterise the CLS amplitudes and $H_2$ as follows
\begin{gather*}
    H_0 = \begin{pmatrix}
       0 & 1 & 0 \\
       1 & 0 & 1 \\
       0 & 1 & 0
   \end{pmatrix}, \,
    \vpsi_1 = \begin{pmatrix}
       a \\
       b \\
       c
   \end{pmatrix}, \, 
   \vpsi_2 = \begin{pmatrix}
       d \\
       e \\
       f
   \end{pmatrix}, \\
    H_2 = Q_{12} \ket{u}\bra{v} Q_{12}, \\
    \vec{u} = (u_1,u_2,u_3),\ \ \vec{v} = (v_1,v_2,v_3).
\end{gather*}
The nonlinear constraints~\eqref{eq:u21-cls-cons-app} yield $b= -\sqrt{2} f, d = f, c = e = 0$. Then it follows that
\begin{equation*}
    \begin{aligned}
    	H_1 &= \begin{pmatrix}
            -\frac{1}{\sqrt{2}} & -\frac{a}{2 f} & -\frac{1}{\sqrt{2}} \\
            \frac{2 f}{a}-\frac{a}{2 f} & \frac{4-\frac{a^2}{f^2}}{2 \sqrt{2}} & \frac{2 f}{a}-\frac{a}{2 f} \\
            \frac{1}{\sqrt{2}} & \frac{a}{2 f} & \frac{1}{\sqrt{2}} \\
        \end{pmatrix},\\ 
        H_2 &= \begin{pmatrix}
            \sqrt{2} A f^2 & a A f & -\sqrt{2} A f^2 \\
            a A f & \frac{a^2 A}{\sqrt{2}} & -a A f \\
            -\sqrt{2} A f^2 & -a A f & \sqrt{2} A f^2 \\
        \end{pmatrix},
    \end{aligned}
\end{equation*}
where 
\begin{gather*}
    A = \frac{\left(\sqrt{2} a u_2+2 f (u_1-u_3)\right) \left(a v_2+2 f (v_1-v_3)\right)}{\left(a^2+4 f^2\right)^2}.
\end{gather*}
The FB energy $\efb$ is then obtained from the first non-linear constraint~\eqref{eq:u21-cls-cons-app}.

The specific lattice structure of the Hamiltonian in Fig.~\ref{fig:nn-u21-example}(a) corresponds to the following choices of free parameter:
\begin{align*}
	x_3 &= x_1, y_3 = y_1, x_2 = 1, y_2 = 5, u_3 = u_1,\\ 
	v_3 &= v_1, u_2 = 1, v_2 = -5, f = -1, a = 1.
\end{align*}
The hopping matrices and CLS amplitudes read
\begin{align*}
    H_1 &= \begin{pmatrix}
        -\frac{1}{\sqrt{2}} & \frac{1}{2} & -\frac{1}{\sqrt{2}} \\
        -\frac{3}{2} & \frac{3}{2 \sqrt{2}} & -\frac{3}{2} \\
        \frac{1}{\sqrt{2}} & -\frac{1}{2} & \frac{1}{\sqrt{2}} \\
    \end{pmatrix},\\ 
    H_2 &= \begin{pmatrix}
        -\frac{2}{5} & \frac{\sqrt{2}}{5} & \frac{2}{5} \\
        \frac{\sqrt{2}}{5} & -\frac{1}{5} & -\frac{\sqrt{2}}{5} \\
        \frac{2}{5} & -\frac{\sqrt{2}}{5} & -\frac{2}{5} \\
    \end{pmatrix}, \\
    \vpsi_1 & = \begin{pmatrix} 1 \\ \sqrt{2} \\ 0 \end{pmatrix},\ \
    \vpsi_2 = \begin{pmatrix} -1 \\ 0 \\ -1 \end{pmatrix}.
\end{align*}

\subsection{The $U=(2,2,1)$ case and $\nu=3$}
\label{app:nn-u221}

The eigenvalue problem and destructive interference conditions in Eq.~\eqref{eq:u22-gen-eig-prob} become
\begin{equation}
    \begin{aligned}
        & H_1 \vpsi_2 + H_2 \vpsi_3 = (\efb -H_0) \vpsi_2, \\
        & H_1^{\dagger} \vpsi_1 = (\efb - H_0) \vpsi_2, \\
        & H_2^{\dagger} \vpsi_1 = (\efb - H_0) \vpsi_3, \\
        & H_1 \vpsi_1 = H_1^{\dagger} \vpsi_2 = H_1 \vpsi_3 = 0, \\
        & H_2 \vpsi_1 = H_2 \vpsi_2 = H_2^{\dagger} \vpsi_3 = 0, \\
        & H_1^\dagger \vpsi_3 + H_2^{\dagger} \vpsi_2 = 0.
    \end{aligned}  
    \label{eq:lshape-gen-eig-prob-app}
\end{equation} 
Using the destructive interference conditions, we eliminate $H_1,H_2$ from the eigenproblem and obtain the nonlinear constraints on the CLS amplitudes:
\begin{equation}
    \begin{aligned}
        & \mel{\psi_2}{H_0}{\psi_1} = \efb \bra{\psi_2}\ket{\psi_1}, \\
        & \mel{\psi_3}{H_0}{\psi_1} = \efb \bra{\psi_3}\ket{\psi_1}, \\
        & \mel{\psi_3}{H_0}{\psi_2} = \efb \bra{\psi_3}\ket{\psi_2}, \\
        & \mel{\psi_2}{\efb - H_0}{\psi_2} + \mel{\psi_3}{\efb - H_0}{\psi_3} \\
        &= \mel{\psi_1}{\efb - H_0}{\psi_1}.
    \end{aligned}
    \label{eq:non-lin-const-L-shape-app}
\end{equation} 

\subsubsection{Linearly independent CLS components} 
\label{app:u221-special}

In this case, as explained in the main text, the destructive interference condition involving both $H_1$ and $H_2$ decouples into two separate conditions. Then the eigenvalue problem Eq.~\eqref{eq:lshape-gen-eig-prob-app} reads
\begin{equation}
    \begin{aligned}
        & H_1 \vpsi_2 + H_2 \vpsi_3 = (\efb -H_0) \vpsi_1, \\
        & H_1^{\dagger} \vpsi_2 = (\efb - H_0) \vpsi_2, \\
        & H_2^{\dagger} \vpsi_2 = (\efb - H_0) \vpsi_3, \\
        & H_1 \vpsi_2 = H_1^{\dagger} \vpsi_2 = H_1 \vpsi_3 = H_1^{\dagger} \vpsi_3 = 0, \\
        & H_2 \vpsi_2 = H_2^{\dagger} \vpsi_3 = H_2 \vpsi_2 = H_2^{\dagger} \vpsi_2 = 0.
    \end{aligned}  
    \label{eq:lshap-eig-prob}
\end{equation} 
The nonlinear constraints on the amplitudes of the CLS are given by the same Eq.~\eqref{eq:non-lin-const-L-shape-app}. Assuming that the nonlinear constraints are resolved, we provide below the solution to Eq.~\eqref{eq:lshap-eig-prob}. We use the following single projector choices $H_1 = Q_{23} \ket{x}\bra{y} Q_{13}$ and $H_2 = Q_{23}\ket{v}\bra{w} Q_{12}$, where the transverse projectors are enforced by the desctructive interference conditions. Plugging in these expression into the eigenproblem~\eqref{eq:lshap-eig-prob} and using the nonlinear constraints we find  the hopping matrices:
\begin{equation}
    \begin{aligned}
        H_1 & = \frac{(\efb - H_0) \ket{\psi_1} \bra{\psi_2} (\efb - H_0)}{\mel{\psi_1}{\efb - H_0}{\psi_1}}, \\
        H_2 & = \frac{(\efb - H_0) \ket{\psi_1} \bra{\psi_3} (\efb - H_0)}{\mel{\psi_1}{\efb - H_0}{\psi_1}}.
    \end{aligned} 
    \label{eq:u221-spe-sol}
\end{equation} 

\subsubsection{Linearly dependent CLS components} 

The linear dependence of the CLS amplitudes $\vpsi_1,\vpsi_2,\vpsi_3$ reads
\begin{equation}
    \ket{\psi_1} = \alpha\ket{\psi_2} + \beta\ket{\psi_3}.
\end{equation} 
This CLS is not necessarily reducible to the $U=1$ class as long as the amplitudes are not \emph{all} mutually proportional: For example, the Lieb lattice falls under the $U=(2,2,1)$ case and has $\vpsi_1=\vpsi_2 + \vpsi_3$. The constraints on the CLS~\eqref{eq:non-lin-const-L-shape-app} in this case become
\begin{equation}
    \begin{aligned}
        \mel{\psi_2}{\efb - H_0}{\psi_2} & =0, \\
        \mel{\psi_3}{\efb - H_0}{\psi_3} & =0, \\
        \mel{\psi_3}{\efb - H_0}{\psi_2} & =0, \\
        (\efb - H_0)(\alpha\vert \psi_2 \rangle + \beta \vert \psi_3 \rangle) &= 0.
    \end{aligned}
    \label{eq:u221-cls-cons-3band}
\end{equation}
For the given $\efb$, $H_0$ and CLS amplitudes, satisfying the above constraints, the eigenvalue problem and destructive interference conditions~\eqref{eq:lshape-gen-eig-prob-app} become
\begin{equation}
    \begin{aligned}
        \beta\bra{\psi_3} H_1 &= \bra{\psi_2} \left(\efb - H_0\right), \\
        \alpha\bra{\psi_2} H_2 &= \bra{\psi_3} \left(\efb - H_0\right), \\
        H_1\ket{\psi_2} &= H_1\ket{\psi_3} = 0, \\
        H_2\ket{\psi_2} &= H_2\ket{\psi_3} = 0, \\
        \bra{\psi_2} H_1 &= 0, \\
        \bra{\psi_3} H_2 &= 0.
    \end{aligned}
    \label{eq:u221-eig-prob-3band}
\end{equation}
These are two decoupled inverse eigenvalue problems for $H_1$ and $H_2$ that we resolve in the same way as before.~\cite{maimaiti2019universal} The hopping matrices read
\begin{equation}
    \begin{aligned}
        H_1 &= \frac{Q_{2}\ket{u}\bra{\psi_2}\left(\efb - H_0\right)}{\mel{\psi_3}{Q_2}{u}}, \\
        H_2 &= \frac{Q_{3}\ket{v}\bra{\psi_3}\left(\efb - H_0\right)}{\mel{\psi_2}{Q_3}{v}},
    \end{aligned}
    \label{eq:u221-hxy-sol-3band}
\end{equation} 
where $\efb,\ H_0, \vec{u}=(u_1,u_2,u_3), \vec{v}=(v_1,v_2,v_3)$ are free parameters; $\vpsi_1 ,\ \vpsi_2 $ are coinstrained by Eq.~\eqref{eq:u221-cls-cons-3band} while $\vec{u}$ should not to be proportional to $\vpsi_2$, and $\vec{v}$ should not be proportional to $\vpsi_3$.

\subsubsection{Three band examples for the linearly dependent case} 
\label{app:nn-u221-eg}

We choose $H_0$ and parameterise the CLS amplitudes as follows
\begin{equation} 
    H_0 = \begin{pmatrix}
        0 & 0 & 0 \\
        0 & 1 & 0 \\
        0 & 0 & \epsilon \\
    \end{pmatrix},
    \quad
    \psi_2 = \begin{pmatrix} a \\ b \\ c \end{pmatrix}, 
    \quad
    \psi_3 = \begin{pmatrix} e \\ f \\ g \end{pmatrix}.
    \label{eq:nu3-ex-psi23-app}
\end{equation}

\paragraph{The Tasaki and Lieb lattice families}

Here we demonstrate how the Lieb lattice with a flatband at $\efb=0$ enforced by the chiral symmetry, and the related Tasaki lattice are derived from our solution. The nonlinear constraints~\eqref{eq:u221-cls-cons-3band} give
\begin{equation*}
    \psi_2 = \begin{pmatrix} a \\ c \sqrt{\epsilon } (\pm i) \\ c \\ \end{pmatrix},
    \psi_3 = \begin{pmatrix} e \\ g \sqrt{\epsilon } (\pm i) \\ g \\ \end{pmatrix}
\end{equation*}
To reproduce the Lieb lattice Hamiltonian the following unitary transformation is used
\begin{equation*}
    H_0^\prime = U H_0 U^\dagger,\ \
    \psi _2^\prime = U \psi_2,\ \ \ 
    \psi _3^\prime = U \psi_3
\end{equation*}
where 
\begin{widetext}
    \begin{equation*}
        U = \begin{pmatrix}
            \cos (\theta ) \cos (\varphi ) & \cos (\theta ) \sin (\varphi ) \sin (\phi )-\sin (\theta ) \cos (\phi ) & \cos (\theta ) \sin (\varphi ) \cos (\phi )+\sin (\theta ) \sin (\phi ) \\
            \sin (\theta ) \cos (\varphi ) & \sin (\theta ) \sin (\varphi ) \sin (\phi )+\cos (\theta ) \cos (\phi ) & \sin (\theta ) \sin (\varphi ) \cos (\phi )-\cos (\theta ) \sin (\phi ) \\
            -\sin (\varphi ) & \cos (\varphi ) \sin (\phi ) & \cos (\varphi ) \cos (\phi ) \\
        \end{pmatrix}
    \end{equation*}
\end{widetext}
The Lieb lattice is recovered for the following choices of parameters:
\begin{equation*}
    \begin{aligned}
        b &= i \frac{a\sqrt{\epsilon }}{\sqrt{2}},\ \  f= -\frac{i a \alpha  \sqrt{\epsilon }}{\beta  \sqrt{2}},\ \ g= -\frac{a \alpha }{\beta  \sqrt{2}}, \\
        c &= \frac{a}{\sqrt{2}},\ \ e= a\ \ \theta = -\frac{\pi }{2},\ \ \ \phi = \frac{3 \pi }{4}, \\ 
        \varphi &=  -\frac{\pi }{4},\ \ \epsilon = -1,\ \ \ c= \frac{a}{\sqrt{2}},\ \ \beta =  \alpha.
    \end{aligned}
\end{equation*}
Then the CLS amplitudes are given by
\begin{equation*}
    \begin{aligned}
        \psi_2^{\prime} &= \begin{pmatrix} 0 \\ -\sqrt{2}a \\ 0 \\ \end{pmatrix}, 
        \psi_3^{\prime} = \begin{pmatrix} 0 \\ 0 \\ \sqrt{2}a  \\ \end{pmatrix},
        \psi_1^{\prime} &= \alpha \begin{pmatrix} 0 \\ -\sqrt{2}a \\ \sqrt{2}a \\ \end{pmatrix}.
    \end{aligned}
\end{equation*}
With the choice $\theta =-\frac{\pi }{2},\phi =\frac{3 \pi }{4},\epsilon =-1$ it follows 
\begin{equation*}
    \begin{aligned}
        H_0^{\prime} & =\begin{pmatrix}
            0 & \sin(\varphi ) & \cos(\varphi ) \\
            \sin(\varphi) & 0 & 0 \\
            \cos(\varphi) & 0 & 0 \\
        \end{pmatrix}.
    \end{aligned}
\end{equation*}
The hopping matrices $H_1^{\prime}, H_2^{\prime}$ are given by Eq.~\eqref{eq:u221-hxy-sol-3band}:
\begin{equation*}
    H_1^{\prime} = \begin{pmatrix}
        -\frac{u_1 \sin(\varphi ) }{u_3 \alpha } & 0 & 0 \\
        0 & 0 & 0 \\
        -\frac{\sin (\varphi ) }{\alpha } & 0 & 0 \\
    \end{pmatrix},
    H_2^{\prime} = \begin{pmatrix}
        \frac{v_1 \cos (\varphi )}{v_2 \alpha } & 0 & 0 \\
        \frac{\cos (\varphi )}{\alpha } & 0 & 0 \\
        0 & 0 & 0 \\
    \end{pmatrix}
\end{equation*} 
The above hopping matrices correspond to the family of Tasaki lattices.~\cite{tasaki1992ferromagnetism} For $u_1=v_1=0$, we retrieve the hopping matrices of the Lieb lattice family.

\paragraph{Obtaining the example in Fig.~\ref{fig:nn-u221-example}(a,b)}
\label{app:u221-eg-gen} 

We choose $\efb = \epsilon$, $\epsilon\neq 0, 1$ and find
\begin{equation*}
    a = -\frac{b \sqrt{-f^2 (\epsilon -1)}}{f \sqrt{\epsilon }}, \,
    e = -\frac{\sqrt{f^2-f^2 \epsilon }}{\sqrt{\epsilon }}, \,
   	\beta = -\frac{\alpha b}{f}.
\end{equation*}
Then Eq.~\eqref{eq:u221-hxy-sol-3band} yields the following solution
\begin{equation*}
	\begin{aligned} 
		\psi_2 &= \begin{pmatrix}
		    -\frac{b D}{f \sqrt{\epsilon }} \\
		    b \\
		    c
		\end{pmatrix},\ \
		\psi_3= 
		\begin{pmatrix}
		    -\frac{D}{\sqrt{\epsilon }} \\
		    f \\
		    g
		\end{pmatrix}, \ \
		\psi_1 = \alpha \psi_2 + \beta \psi_3, \\
		H_1 &= \begin{pmatrix}
            \frac{b B D \epsilon }{A f} & -\frac{b B (\epsilon -1) \sqrt{\epsilon }}{A} & 0 \\
            \frac{b C D \sqrt{\epsilon }}{A f} & \frac{b C (1-\epsilon )}{A} & 0 \\
            \frac{b^2 D \sqrt{\epsilon }}{\alpha  c f^2-\alpha  b f g} & \frac{b^2-b^2 \epsilon }{\alpha  c f-\alpha  b g} & 0 \\
		\end{pmatrix}, \\ 
		H_2 &= \begin{pmatrix}
            -\frac{D f G \epsilon }{F} & \frac{f^2 G (\epsilon -1) \sqrt{\epsilon }}{F} & 0 \\
            -\frac{D f H \sqrt{\epsilon }}{F} & -\frac{f^2 H (1-\epsilon )}{F} & 0 \\
     		-\frac{D f \sqrt{\epsilon }}{\alpha  c f-\alpha  b g} & \frac{f^2 (\epsilon -1)}{\alpha  (c f-b g)} & 0 \\
		\end{pmatrix},
	\end{aligned}
\end{equation*}
where 
\begin{equation*}
    \begin{aligned}
        D &= \sqrt{-f^2 (\epsilon -1)},\\ 
        A &= \alpha  (c f-b g) \left(b f u_3+c D u_1 \sqrt{\epsilon }-c f u_2 \epsilon \right), \\ 
        B &= b^2 \left(D u_2+f u_1 \sqrt{\epsilon }\right)+b c D u_3+c^2 f u_1 \sqrt{\epsilon }, \\ 
        C &= b^2 \left(D u_1 \sqrt{\epsilon }+f u_2 (1-\epsilon )\right)-b c f u_3 \epsilon +c^2 f u_2 \epsilon, \\ 
        F &= \alpha  (c f-b g) \left(D g v_1 \sqrt{\epsilon }+f^2 v_3-f g v_2 \epsilon \right), \\ 
        G &= D f v_2+g \left(D v_3+g v_1 \sqrt{\epsilon }\right)+f^2 v_1 \sqrt{\epsilon }, \\ 
        H &= D f v_1 \sqrt{\epsilon }+f^2 (v_2-v_2 \epsilon )-f g v_3 \epsilon +g^2 v_2 \epsilon .
    \end{aligned}
\end{equation*}
We use the following parameter values
\begin{equation*}
    \begin{split}
        u_1 &= 0,u_2= 0,u_3= 2,\alpha = 1,
        \beta = 1,v_1= 0,v_2= 0,\\ 
        v_3 &= 1,  a = 0,b= -1,g= 0, 
        \epsilon = \frac{1}{2},c= 2,f= -1, 
    \end{split}
\end{equation*} 
finding the CLS and hopping matrices
\begin{equation*}
	\begin{aligned}
		\psi_2 = \begin{pmatrix} -1 \\ -1 \\ 2 \end{pmatrix},\ \
		\psi_3 = \begin{pmatrix} -1 \\ -1 \\ 0 \end{pmatrix},\ \
		\psi_1 = \begin{pmatrix} -2 \\ -2 \\ 2 \end{pmatrix}, \\
		H_0 = \begin{pmatrix}
            0 & 0 & 0 \\
     		0 & 1 & 0 \\
     		0 & 0 & \frac{1}{2} \\
		\end{pmatrix},
		H_1 = \begin{pmatrix}
            \frac{1}{4} & -\frac{1}{4} & 0 \\
     		\frac{1}{4} & -\frac{1}{4} & 0 \\
            \frac{1}{4} & -\frac{1}{4} & 0 \\
		\end{pmatrix}, 
		H_2 = \begin{pmatrix}
            0 & 0 & 0 \\
     		0 & 0 & 0 \\
            -\frac{1}{4} & \frac{1}{4} & 0 \\
		\end{pmatrix}
	\end{aligned}
\end{equation*}
The lattice structure and band structure corresponding to the above hopping matrices is shown in Fig.~\ref{fig:nn-u221-example}(a,b).

\subsection{The $U=(2,2,0)$ case with three bands}
\label{app:nn-u220-nu3} 

Putting the values $U_2=2,\ s=0$ to Eq.~\eqref{eq:u22-gen-eig-prob}, we find the following eigenvalue problem
\begin{equation}
    \begin{aligned}
        H_1 \psi_{2,1} + H_2 \psi_{1,2} &= \left(\efb - H_0\right) \psi_{1,1}\\
        H_1^{\dagger} \psi_{1,1} + H_2 \psi_{2,2} &= \left(\efb - H_0\right) \psi_{2,1},\\
        H_1 \psi_{2,2} + H_2^{\dagger} \psi_{1,1} &= \left(\efb - H_0\right) \psi_{1,2},\\
        H_1^{\dagger} \psi_{1,2} + H_2^{\dagger} \psi_{2,1} &= \left(\efb - H_0\right) \psi_{2,2},
    \end{aligned}
    \label{eq:square_u22_eig_prob}
\end{equation}
and destructive interference conditions
\begin{equation}
    \begin{aligned}
        H_1 \psi_{1,1} &= H_1 \psi_{1,2} = H_2 \psi_{1,1 } = H_2 \psi_{2,1} = 0,\\ 
        H_1^{\dagger} \psi_{2,1 } &= H_1^{\dagger} \psi_{2,2} = H_2^{\dagger} \psi_{1,2} = H_2^{\dagger} \psi_{2,2} = 0.
    \end{aligned}
    \label{eq:square_u22_di}
\end{equation}
We assume that $H_1, H_2$ have two zero modes and parameterize the hopping matrices $H_1,H_2$ in the following way: 
\begin{gather}
    H_1 = \ket{x}\bra{y} = \begin{pmatrix}
        a d & a e & a f \\
        b d & b e & b f \\
        c d & c e & c f \\
    \end{pmatrix}, \\ 
    H_2 = \ket{u}\bra{v} = \begin{pmatrix}
        g r & g s & g t \\
        h r & h s & h t \\
        l r & l s & l t \\
    \end{pmatrix},
    \label{eq:Hxy}
\end{gather} 
where
\begin{equation*}
    \ket{x} = \begin{pmatrix} a \\ b \\ c \\ \end{pmatrix},
    \ket{y} = \begin{pmatrix} d \\ e \\ f \\ \end{pmatrix},
    \ket{u} = \begin{pmatrix} g \\ h \\ l \\ \end{pmatrix},
    \ket{v} = \begin{pmatrix} r \\ s \\ t \\ \end{pmatrix}.
\end{equation*}
The zero more of $H_1,H_2$ are given by (the top two lines correspond to $H_1$, the bottom two -- to $H_2$; in every raw the first two elements are the right eigenvectors, while the last two are the left ones)
\begin{equation*}
    \begin{aligned}
        \begin{pmatrix} -f \\ 0 \\ d \\ \end{pmatrix},\begin{pmatrix} -e \\ d \\ 0 \\ \end{pmatrix}, 
        \begin{pmatrix} -c \\ 0 \\ a \\ \end{pmatrix},\begin{pmatrix} -b \\ a \\ 0 \\ \end{pmatrix},\\
        \begin{pmatrix} -t \\ 0 \\ r \\ \end{pmatrix},\begin{pmatrix} -s \\ r \\ 0 \\ \end{pmatrix},
        \begin{pmatrix} -l \\ 0 \\ g \\ \end{pmatrix},\begin{pmatrix} -h \\ g \\ 0 \\ \end{pmatrix}.
    \end{aligned}
\end{equation*}
Next we enforce the constraints on $H_1,H_2$ and the CLS amplitudes by the destructive interference conditions~\eqref{eq:square_u22_di}. Since $\psi _{1,1}$ is the right zero eigenmode of both $H_1$ and $H_2$, it has to be perpendicular to both $\vec{y}$ and $\vec{v}$, or equivalently it is parallel to the cross product of $\vec{y}$ and $\vec{v}$ and also parallel to one of the right zero eigenvectors of $H_1,H_2$:
\begin{equation*}
    \vpsi_{1,1} = \alpha (y \times v) \parallel
    \begin{pmatrix} -f \\ 0 \\ d \\ \end{pmatrix}
    \parallel
    \begin{pmatrix} -t \\ 0 \\ r \\ \end{pmatrix}. 
\end{equation*}
where we have introduced the proportionality factor $\alpha$, that we set to $1$ for convenience. Treating the other CLS amplitudes in the same way (and setting the proportionality factors to $1$ as well) we arrive at the following set of constraints on the elements of the CLS amplitudes:
\begin{equation}
    \begin{aligned}
        t &= f, b = s, e = h,c = l,\\
        d &= r = a = g.
    \end{aligned}
    \label{eq:Hxy_elements}
\end{equation}
Then the expressions for all $\psi$ reduce to the following equations:
\begin{equation}
    \begin{aligned}
        \psi_1 &= \begin{pmatrix}
            (-b f+e f) \alpha  \\
            0 \\ 
            (a b-a e) \alpha \} \\
        \end{pmatrix} = \alpha^\prime
        \begin{pmatrix}
            -f \\ 
            0 \\ 
            a 
        \end{pmatrix},\\
        \psi_2 &= \begin{pmatrix}
            (-b c+b f) \beta  \\
            (a c-a f) \beta  \\
            0 \\
        \end{pmatrix} = \beta^\prime
        \begin{pmatrix}
            -b  \\
            a  \\
            0 
        \end{pmatrix},\\
        \psi_3 &= \begin{pmatrix}
            (c e-e f) \gamma  \\
            (-a c+a f) \gamma  \\
            0 \\
        \end{pmatrix} = \gamma^\prime
        \begin{pmatrix}
            -e  \\
            a  \\
            0 \\
        \end{pmatrix},\\
        \psi_4 &= \begin{pmatrix}
            (b c-c e) \eta  \\
            0 \\
            (-a b+a e) \eta  \\
        \end{pmatrix} = \eta^\prime
        \begin{pmatrix}
            -c  \\
            0 \\
            a  \\
        \end{pmatrix}.
    \end{aligned}
    \label{eq:psies}
\end{equation}
where $\alpha^\prime, \beta^\prime, \gamma^\prime, \eta^\prime$ are given by
\begin{equation}
	\begin{aligned}
    	\alpha^\prime &=(b-e) \alpha ,\ \ \beta^\prime=(c-f) \beta ,\\
    	\gamma^\prime &=-c \gamma + f \gamma ,\ \ \eta^\prime=-b \eta +e \eta. 
    \end{aligned}
    \label{eq:psies_coef}
\end{equation}

We can set one of the pre-factors to $1$: we choose $\eta =1$. Then Eq.~\eqref{eq:psies} becomes
\begin{equation*}
    \begin{aligned}
        \psi_1 &=
        \begin{pmatrix}
            -(b-e) f \alpha \\
            0 \\
            a (b-e) \alpha \\
        \end{pmatrix},
        \psi_2 =
        \begin{pmatrix}
            -b (c-f) \beta  \\
            a (c-f) \beta  \\
            0 \\
        \end{pmatrix},\\
        \psi_3 &=
        \begin{pmatrix}
            -e (-c \gamma +f \gamma ) \\
            a (-c \gamma +f \gamma ) \\
            0 \\
        \end{pmatrix},
        \psi_4 =
        \begin{pmatrix}
            -c (-b+e) \\
            0 \\
            a (-b+e) \\
        \end{pmatrix}.
    \end{aligned}
\end{equation*}
We choose $H_0$ as
\begin{equation*}
    H_0 = \begin{pmatrix}
        0 & 0 & 0 \\
        0 & 1 & 0 \\
        0 & 0 & \epsilon  \\
    \end{pmatrix}.
\end{equation*}
Putting Eqs.~\eqref{eq:Hxy_elements} and~\eqref{eq:psies_coef} into Eq.~\eqref{eq:Hxy}, we get the hopping matrices
\begin{gather}
    H_1 = \begin{pmatrix}
        a^2 & a e & a f \\
        a b & b e & b f \\
        a c & c e & c f \\
    \end{pmatrix},
    H_2 =\begin{pmatrix}
        a^2 & a b & a f \\
        a e & b e & e f \\
        a c & b c & c f \\
    \end{pmatrix},
    \label{eq:Hxy_new}
\end{gather}
The eigenvalue problem~\eqref{eq:square_u22_eig_prob} becomes
\begin{widetext}
    \begin{equation*}
        \begin{aligned}
            \begin{pmatrix}
                (b-e) \left(-a^2 (c-f) (\beta +\gamma )+f \alpha  \efb \right) \\
                -a (b-e) (c-f) (b \beta +e \gamma ) \\
                a (b-e) (-c (c-f) (\beta +\gamma )+\alpha  (\epsilon -\efb )) \\
            \end{pmatrix}
            & = 0,\\
            \begin{pmatrix}
                (c-f) \left(a^2 (b-e) (1+\alpha )+b \beta  \efb \right) \\
                a (c-f) ((b-e) e (1+\alpha )+\beta -\beta  \efb ) \\
                a (b-e) (c-f) (c+f \alpha ) \\
            \end{pmatrix} &= 0,\\
            \begin{pmatrix}
                (c-f) \left(a^2 (b-e) (1+\alpha )-e \gamma  \efb \right) \\
                a (c-f) (b (b-e) (1+\alpha )+\gamma  (-1+\efb )) \\
                a (b-e) (c-f) (c+f \alpha ) \\
            \end{pmatrix} &= 0, \\
            \begin{pmatrix}
                (b-e) \left(-a^2 (c-f) (\beta +\gamma )-c \efb \right) \\
                -a (b-e) (c-f) (b \beta +e \gamma ) \\
                a (b-e) (-(c-f) f (\beta +\gamma )-\epsilon +\efb ) \\
            \end{pmatrix} &= 0.
        \end{aligned}
    \end{equation*}
\end{widetext}
Assuming that $a\neq 0,\ c\neq f,\ b\neq e$ we solve the above equations (according to Eqs.~(\ref{eq:psies}-\ref{eq:psies_coef}) $a=0$ makes the $\psi_{i=1,\dots,4}$ proportional, i.e. we find $U=1$ CLS; while $c\neq f,\ b\neq e$ enforces $\psi_{i=1,\dots,4}=0$). Setting $a=1$ for convenience the solution is
\begin{widetext}
    \begin{equation*}
        \begin{aligned}
            b & =\frac{\text{sgn}(\efb)\sqrt{2} (\efb - 1)\vert 1 + \alpha\vert}{\sqrt{-\sqrt{-\alpha (1 - \efb )^2 \efb ^4 \left(4 (1+\alpha )^2 - \alpha \efb ^2\right)} + (\efb - 1) \efb  \left(\alpha \efb^2 -2 (1+\alpha)^2 \right)}}, \\
            c &= \frac{\sqrt{\alpha(\efb -\epsilon}}{\sqrt{\efb }}, \quad f = -\frac{\sqrt{\efb - \epsilon}}{\sqrt{\alpha\efb}}, \\
             e &=-\frac{|1+\alpha|^{-1}}{\efb\sqrt{2}}\sqrt{-\sqrt{-\alpha (\efb - 1)^2 \efb ^4 \left(4 (1+\alpha )^2 - \alpha \efb ^2\right)} + (\efb - 1) \efb \left(\alpha \efb ^2 - 2(1+\alpha)^2\right)}, \\
            \beta &= \frac{-\alpha (\efb - 1) \efb ^3+\sqrt{-\alpha (1 - \efb )^2 \efb ^4 \left(4 (1+\alpha )^2 - \alpha \efb ^2\right)}}{2 (1+\alpha ) (\efb - 1) \efb ^2},\\ \ \ \
            \gamma &= -\frac{\alpha (\efb - 1) \efb ^3 +\ sqrt{-\alpha (\efb - 1)^2 \efb ^4 \left(4 (1+\alpha )^2 - \alpha  \efb ^2\right)}}{2 (1+\alpha ) (\efb - 1) \efb ^2}.
        \end{aligned}
        \label{eq:u22-nn-sol-app}
    \end{equation*}
\end{widetext}
Pluging these solutions into Eq.~\eqref{eq:Hxy_new} we get the follwing hopping matrices
\begin{equation*}
        H_1 = 
        \begin{pmatrix}
            1 & -\frac{B}{\sqrt{2}} & -\frac{\sqrt{\efb -\epsilon }}{\sqrt{\alpha } \sqrt{\efb }} \\
            \frac{\sqrt{2} (\efb -1)}{B \efb } & \frac{1}{\efb }-1 & -\frac{\sqrt{2} (\efb -1) \sqrt{\efb -\epsilon }}{\sqrt{\alpha } B \efb ^{3/2}} \\
            \frac{\sqrt{\alpha } \sqrt{\efb -\epsilon }}{\sqrt{\efb }} & -\frac{\sqrt{\alpha } B \sqrt{\efb -\epsilon }}{\sqrt{2} \sqrt{\efb }} & \frac{\epsilon }{\efb }-1 \\
        \end{pmatrix},
\end{equation*}
\begin{equation*}
        H_2 = 
        \begin{pmatrix}
            1 & \frac{\sqrt{2} (\efb -1)}{B \efb } & -\frac{\sqrt{\efb -\epsilon }}{\sqrt{\alpha } \sqrt{\efb }} \\
            -\frac{B}{\sqrt{2}} & \frac{1}{\efb }-1 & \frac{B \sqrt{\efb -\epsilon }}{\sqrt{2} \sqrt{\alpha } \sqrt{\efb }} \\
            \frac{\sqrt{\alpha } \sqrt{\efb -\epsilon }}{\sqrt{\efb }} & \frac{\sqrt{2} \sqrt{\alpha } (\efb -1) \sqrt{\efb -\epsilon }}{B \efb ^{3/2}} & \frac{\epsilon }{\efb }-1 \\
        \end{pmatrix}
\end{equation*}
where 
\begin{align*}
    A &= \sqrt{-\alpha  (\efb -1)^2 \efb ^4 \left(4 (\alpha +1)^2-\alpha  \efb ^2\right)},\\ 
    B &= \sqrt{\frac{(\efb -1) \efb  \left(\alpha  \efb ^2-2 (\alpha +1)^2\right)-A}{(\alpha +1)^2 \efb ^2}}
\end{align*}

This solution has three free parameters $\epsilon, \efb, \alpha$. We choose the following values $\epsilon = -1,\efb = -4,\alpha = 1$ to generate the example shown in Fig.~\ref{fig:nn-u220-example}: $A=0$, $B = \sqrt{5/2}$,
\begin{equation*} 
    \begin{aligned}
        H_0 &= \begin{pmatrix}
            0 & 0 & 0 \\
            0 & 1 & 0 \\
            0 & 0 & -1 \\
        \end{pmatrix}, \\
        H_1 &= \begin{pmatrix}
            1 & -\frac{\sqrt{5}}{2} & -\frac{\sqrt{3}}{2} \\
            \frac{\sqrt{5}}{2} & -\frac{5}{4} & -\frac{\sqrt{15}}{4} \\
            \frac{\sqrt{3}}{2} & -\frac{\sqrt{15}}{4} & -\frac{3}{4} \\
        \end{pmatrix}, \\
        H_2 &= \begin{pmatrix}
            1 & \frac{\sqrt{5}}{2} & -\frac{\sqrt{3}}{2} \\
            -\frac{\sqrt{5}}{2} & -\frac{5}{4} & \frac{\sqrt{15}}{4} \\
            \frac{\sqrt{3}}{2} & \frac{\sqrt{15}}{4} & -\frac{3}{4} \\
        \end{pmatrix}
    \end{aligned}
\end{equation*}

\section{FB generation for three hopping matrices}
\label{app:2d-nnn}

\subsection{$U=(2,1)$} 
\label{app:nnn-u21}

This case is shown in Fig.~\ref{fig:u22-nnn-config}(b). The eigenvalue problem and destructive interference conditions for $H_1$ Eq.~\eqref{eq:u22-nnn-eig-prob} are identical to the case of two hopping matrices and are solved in Appendix~\ref{app:nn-u21}. The only difference are the destructive interference conditions for $H_{2,3}$:
\begin{equation}
    \begin{aligned}
       H_2 \kpsi{1} = H_3 \kpsi{1} & = 0\\
       \bpsi{2} H_2 = \bpsi{2} H_3 & = 0 , \\
       H_2 \kpsi{2} + H_3^\dagger \kpsi{1} &= 0, \\
       \bpsi{1} H_2 + \bpsi{2} H_3^\dagger &= 0. \\
    \end{aligned}
    \label{eq:u21-nnn-eig-prob}
\end{equation}
We define:
\begin{gather*}
    H_2 = Q_2 M_2 Q_3,\quad H_3 = Q_2 M_3 Q_1, \\
    \kx = Q_2\kpsi{1}, \quad\ky = Q_1\kpsi{2}.
\end{gather*}
Then the last two equations~\eqref{eq:u21-nnn-eig-prob} become:
\begin{gather*}
    Q_1 M_3^\dagger Q_2\kx = -Q_2 M_2 Q_1\ky = -Q_{12}\ket{a},\\
    \by Q_2 M_3^\dagger Q_1 = -\bx Q_2 M_2 Q_1 = -Q_{12}\bra{b},
\end{gather*}
where we have introduced two arbitrary vectors $\vec{a}$ and $\vec{b}$. For two bands $\nu=2$ the above equations imply that the last two conditions in~\eqref{eq:u21-nnn-eig-prob} decouple, and therefore the problem reduces to $U=1$ as in the case of two hopping matrices. For the number of bands $\nu>2$ the problem of finding $H_2$ and $H_3$ reduces to two independent inverse eigenvalues problems: one for $M_2$ ($H_2$)
\begin{gather*}
    Q_2 M_2 Q_1\ky = Q_{12}\ket{a},\\
    \bx Q_2 M_2 Q_1 = Q_{12}\bra{b},
\end{gather*}
and a similar problem for the matrix $M_3$ ($H_3$). This is a linear problem: we search for a particular solution as
\begin{gather*}
    Q_2 M_2 Q_1 = \ket{u}\bra{\overline{y}} + \ket{\overline{x}}\bra{v},
\end{gather*}
where we choose the overlined vectors so that: $\bra{\overline{x}}\propto\bra{x}$, $\bra{\overline{x}}\ket{x} = 1$ and $\bra{\overline{y}}\propto\bra{y}$, $\bra{\overline{y}}\ket{y} = 1$. We also assume that $\vec{u}\perp\vec{x}$ and $\vec{v}\perp\vec{y}$. We find upon substitution of the ansatz into the inverse problem:
\begin{gather*}
    Q_2\ket{u} = Q_{12}\ket{a}, \quad \bra{v}Q_1 = \bra{b} Q_{12}.
\end{gather*}
These $\vec{u}$ and $\vec{v}$ have the assumed previously orthogonality properties. It then follows that the full solution  of~\eqref{eq:u21-nnn-m2} is given by:
\begin{gather*}
    H_2 = Q_{12}\ket{a}\bra{\overline{y}}Q_1 + Q_2\ket{\overline{x}}\bra{b}Q_{12} + Q_{12} K_2 Q_{12}.
\end{gather*}
The inverse problem for $H_3$ is resolved the same way with minimal modifications.

\subsection{$U=(2,2,1)$ and $\nu=2$}
\label{app:nnn-u221}

We choose the following $H_0$ and parameterise the CLS amplitudes as follows
\begin{equation*}
    H_0 = 
	\begin{pmatrix}
	    0 & 1 \\
	    1 & 0
    \end{pmatrix}, \,
    \vpsi_1 = \begin{pmatrix} p \\ r \end{pmatrix}, \,
    \vpsi_2 = \begin{pmatrix} s \\ t \end{pmatrix}, \,
    \vpsi_3 = \begin{pmatrix} u \\ v \end{pmatrix}.
\end{equation*}
We parameterize the hopping matrices and solve the eigenvalue problem and destructive interference conditions in Eq.~\eqref{eq:nnn-triang-eig-prob}:
\begin{equation*}
    H_1 = \begin{pmatrix}
        a & \frac{a c}{b} \\
        b & c
    \end{pmatrix}, \,
    H_2 = \begin{pmatrix}
        d & e \\
        \frac{d f}{e} & f
    \end{pmatrix}, \,
    H_3 = \begin{pmatrix}
        g & \frac{g l}{h} \\
        h & l
    \end{pmatrix}.
\end{equation*}
Then one of the possible solutions of the eigenvalue problem~\eqref{eq:nnn-triang-eig-prob} reads
\begin{equation*}
    \begin{aligned}
	    a &= \frac{r (b p-s)}{s^2},\ \ c= -\frac{b p}{r}, \\ d &= \frac{i \sqrt{b^2 \left(p^2+s^2\right) \left(r^2-s^2\right)-2 b p r^2 s+r^2 s^2}}{s^2},\ \  f= 0, \\
    	g &= -\frac{i (b p-s) \sqrt{b^2 \left(p^2+s^2\right) \left(r^2-s^2\right)-2 b p r^2 s+r^2 s^2}}{b s^3}, \\ 
    	l &= u = 0,\\ 
    	h&= -\frac{i \sqrt{b^2 \left(p^2+s^2\right) \left(r^2-s^2\right)-2 b p r^2 s+r^2 s^2}}{r s},\\ 
    	v &= -\frac{i \sqrt{b^2 \left(p^2+s^2\right) \left(r^2-s^2\right)-2 b p r^2 s+r^2 s^2}}{b s}, \\ 
    	e&= -\frac{i p \sqrt{b^2 \left(p^2+s^2\right) \left(r^2-s^2\right)-2 b p r^2 s+r^2 s^2}}{r s^2},\\ 
    	t &= r \left(\frac{1}{b}-\frac{p}{s}\right),\ \ \efb = \frac{b \left(p^2+s^2\right)}{r s}.
    \end{aligned}
\end{equation*}
The corresponding hopping matrices become
\begin{equation*}
    \begin{aligned}
    	H_1 &= \begin{pmatrix}
    	    \frac{r (b p-s)}{s^2} & -\frac{p (b p-s)}{s^2} \\
    	    b & -\frac{b p}{r}
    	\end{pmatrix}, \\
    	H_2 = \begin{pmatrix}
    	    \frac{i A}{s^2} & -\frac{i p A}{r s^2} \\
    	    0 & 0
    	\end{pmatrix}, \,
    	H_3 &= \begin{pmatrix}
    	    -\frac{i (b p-s) A}{b s^3} & 0 \\
    	    -\frac{i A}{r s} & 0
    	\end{pmatrix}, \\
    	\vpsi_1 = \begin{pmatrix} p \\ r \end{pmatrix}, \,
    	\vpsi_2 &= \begin{pmatrix} s \\ r \left(\frac{1}{b}-\frac{p}{s}\right) \end{pmatrix}, \,
    	\vpsi_3 = \begin{pmatrix} 0 \\ - \frac{i A}{b s} \end{pmatrix},
    \end{aligned}
\end{equation*}
where 
\begin{gather*}
    A = \sqrt{b^2 \left(p^2+s^2\right) \left(r^2-s^2\right)-2 b p r^2 s+r^2 s^2} 
\end{gather*}
The network shown in Fig.~\ref{fig:nn-u221-example}(c,d) corresponds to the values $p=1, r=1, b=2, s=1$ and
\begin{align*}
    H_1 &= \begin{pmatrix}
        1 & -1 \\
        2 & -2 \\
    \end{pmatrix}, \ \ 
    H_1 = \begin{pmatrix}
        -\sqrt{3} & \sqrt{3} \\
        0 & 0 \\
    \end{pmatrix}, \\ 
    H_3 &= \begin{pmatrix}
        \frac{\sqrt{3}}{2} & 0 \\
        \sqrt{3} & 0 \\
    \end{pmatrix}, \\
    \psi_1 &= \begin{pmatrix} 1 \\ 1 \end{pmatrix}, \ \ 
    \psi_2 = \begin{pmatrix} 1 \\ - \frac{1}{2} \end{pmatrix}, \ \ 
    \psi_3 = \begin{pmatrix} 0 \\ \frac{\sqrt{3}}{2} \end{pmatrix}
\end{align*}

It is straightforward to check that the well known tight-binding checkerboard lattice is included in the above solution. We choose $H_0$ and the CLS amplitudes of the checkerboard model as input
\begin{equation*}
	H_0 = \begin{pmatrix}
	    0 & b \\
	    b & 0
	\end{pmatrix}, \ \
	\psi_1 = \begin{pmatrix} a \\ -a \end{pmatrix}, \ \
	\psi_2= \begin{pmatrix} a \\ 0 \end{pmatrix}, \ \
	\psi_3 = \begin{pmatrix} 0 \\ -a \end{pmatrix} .
\end{equation*}
The solution yields the checkerboard hopping matrices
\begin{equation*}
	H_1 = \begin{pmatrix}
	    0 & 0 \\
	    b & b
	\end{pmatrix}, \ \
	H_2 = \begin{pmatrix}
	    b & b \\
	    0 & 0
	\end{pmatrix}, \ \
	H_3 = \begin{pmatrix}
	    0 & b \\
	    0 & 0
	\end{pmatrix}.
\end{equation*}

\bibliography{flatband,mbl,frustration}

\end{document}